\documentclass[preprint,review,12pt]{elsarticle}

\usepackage{amssymb}

\usepackage{float}
\usepackage{hyperref}

\biboptions{sort&compress}

\journal{ar$\chi$iv}

\begin{document}

\begin{frontmatter}



\title{Noise2Void for Denoising Atomic Resolution Scanning Transmission Electron Microscopy Images}


\author[inst1]{William Thornley}
\author[inst1,inst2]{Sam Sullivan-Allsop}
\author[inst1]{Rongsheng Cai}
\author[inst1,inst2]{Nick Clark}
\author[inst2,inst3]{Roman Gorbachev}
\author[inst1,inst2]{Sarah J. Haigh}

\affiliation[inst1]{
    organization={Department of Materials},
    addressline={University of Manchester}, 
    postcode={M13 9PL}, 
    country={U.K.}
}
\affiliation[inst2]{
    organization={National Graphene Institute},
    addressline={University of Manchester}, 
    postcode={M13 9PL}, 
    country={U.K.}
}
\affiliation[inst3]{
    organization={Department of Physics and Astronomy},
    addressline={University of Manchester}, 
    postcode={M13 9PL}, 
    country={U.K.}
}

\begin{abstract}

The Noise2Void technique is demonstrated for successful denoising of
atomic-resolution scanning transmission electron microscopy (STEM)
images. The technique is applied to denoising atomic resolution images and
videos of gold adatoms on a graphene surface within a graphene liquid cell,
with the denoised experimental data qualitatively demonstrating improved
visibility of both the Au adatoms and the graphene lattice. The denoising
performance is quantified by comparison to similar simulated data and the
approach is found to significantly outperform both total variation and simple
gaussian blurring. Compared to other denoising methods, the Noise2Void
technique has the combined advantages that it requires no manual intervention
during training or denoising, no prior knowledge of the sample and is
compatible with real time data acquisition rates of at least 45 frames per
second.

\end{abstract}


\begin{highlights}

\item A modified Noise2Void approach is demonstrated for denoising dual-channel bright
field/annular dark field atomic-resolution STEM images
\item The method does not require simulated training data
\item The method is tested for denoising graphene liquid cell STEM images where denoising
improves all metrics by an order of magnitude, a significant improvement compared to total variation
denoising.
\item The dual channel approach is found to reduce training time compared to conventional
mono-channel networks
\item Denoising speed is found to be similar to Gaussian blurring and hence compatible with real
time STEM imaging.

\end{highlights}

\begin{keyword}
deep-learning \sep TEM \sep liquid-cell \sep Noise2Void
\end{keyword}

\end{frontmatter}


\section{Introduction}
\label{sec:introduction}

The latest (scanning) transmission electron microscope ((S)TEM) instruments are capable of spatial resolutions of better than 50 pm, allowing atomic structure to be resolved for many different crystal orientations \cite{Erni2009, Liu2021, Morishita2017, Sawada2009}. Yet imaging at such high magnifications requires high electron fluence to provide sufficient SNR in the resulting images so that information is transferred \cite{Rose1984, Lee2014}. Thus, the limiting factor to achieving atomic resolution (S)TEM imaging of a particular material is often it's stability under a high energy electron beam \cite{Tien2024, Mkhoyan2019, Chen2020}. Image denoising techniques, which aim to improve the signal to noise ratio (SNR) of images after acquisition, have been studied for many years, with a particular focus on removing additive Gaussian white noise (AGWN) \cite{Fan2019}. By improving the SNR, denoising enables information transfer to be retained while lowering the electron fluence (electrons incident per unit area) \cite{deJonge2018, deJonge2019}. Effective denoising therefore has the potential to unlock improved atomic resolution (S)TEM imaging of systems such as metal organic frameworks and pharmaceutical crystals, where spatial information transfer is currently limited by the materials electron beam sensitivity \cite{Tien2024}.

For in situ (S)TEM imaging of dynamic processes, the requirements for excellent electron stability are further increased since the material must survive for multiple image frames \cite{Ilett2020}. Here, successful image denoising provides opportunities for fractionating the material’s critical electron dose across more images: either increasing the number of frames in an image series before damage is observed or increasing frame acquisition rate while retaining information transfer within the individual images.

\begin{figure}[H]
    \centering
    \includegraphics[width=\textwidth]{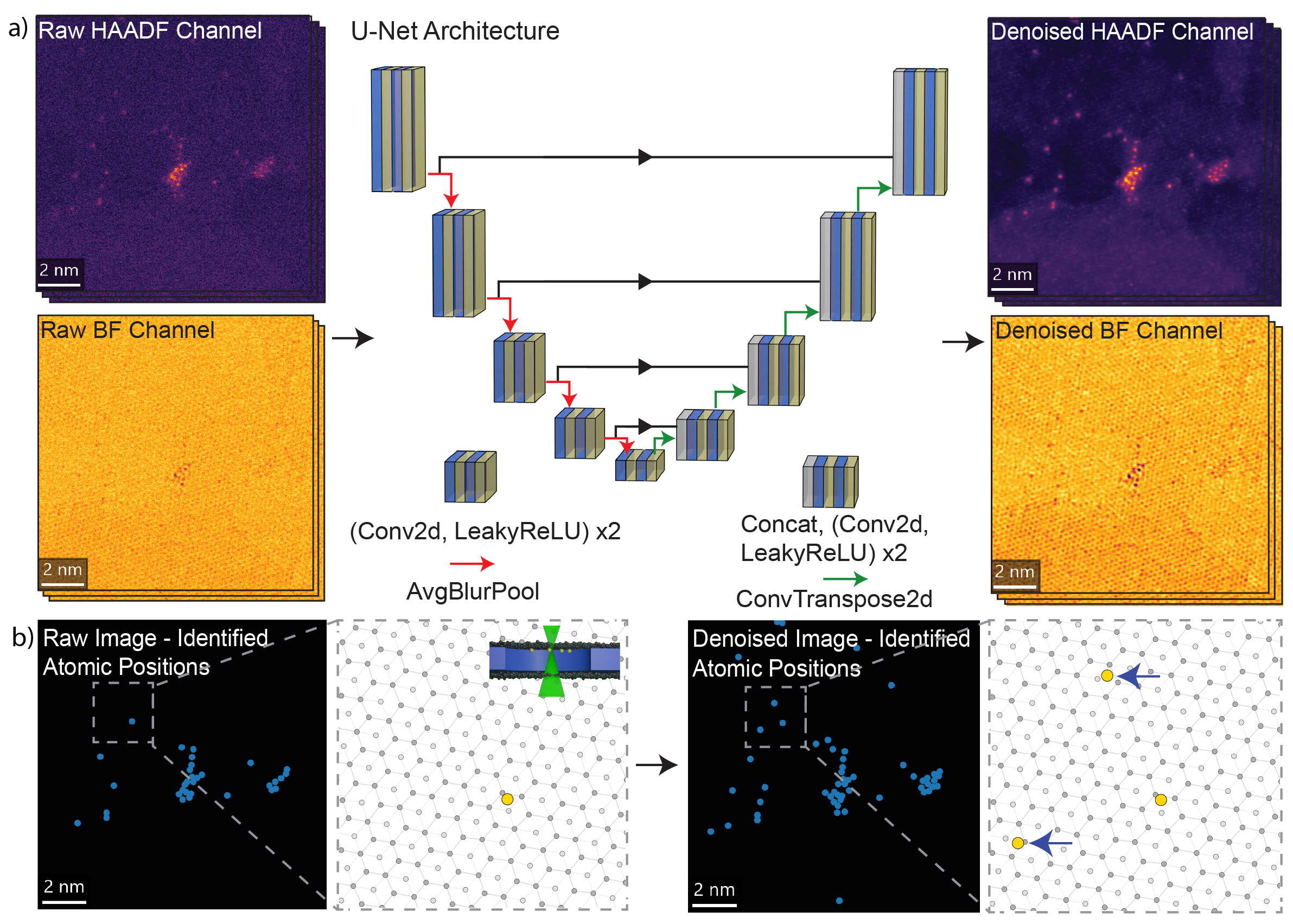}
    \caption{Schematic illustration of the experimental setup and acquired data. (a) An overview of the modified UNet architecture used for Noise2Void denoising where the inputs are noisy dual-channel (HAADF \& BF) STEM video frames (left) and the outputs are the same frames denoised (right). ‘Conv2d’, ‘ConvTranspose2d’ and ‘LeakyReLU’ refer to convolutional blocks, transpose-convolution blocks and leaky ‘ReLU’ activation blocks, respectively, while ‘Concat’ and ‘AvgBlurPool’ refer to channel-wise concatenation and bilinear downsampling \cite{N2V2}, respectively. The black horizontal arrows represent UNet skip connections. (b) The results of a separate atom-finding step applied to the example image above. In the denoised output 89\% more adatom features are identified compared to the raw image, showing the improved performance of atom finding analysis when Noise2Void is used in a data-analysis pipeline. Inset are corresponding atomic diagrams of the magnified region, showing the graphene lattice positions with gold adatoms overlain. The blue arrows highlight gold atoms that were only detected after Noise2Void denoising was applied. Also shown is a model of the TEM liquid cell, showing the convergent electron probe (green), the few-layer graphene windows (black) and the boron nitride spacer layer (blue).}
    \label{fig: cell-schematic}
\end{figure}

In this work we focus on denoising for a particular experimental challenge, one that combines demanding requirements for both spatial and temporal resolution in the STEM. This is the investigation of local atomic motion at solid-liquid interfaces; behaviour that underpins many physical processes such as wetting, adhesion, chemical etching and dissolution and solidification \cite{Howe2004, Zhang2024, Pu2020}. Such studies have recently become achievable in both TEM and STEM imaging modes using advanced liquid cells \cite{Clark2022, Klie2014}. Imaging of volatile liquid-phase samples inside the vacuum of the TEM requires confining a thin volume between two electron transparent windows that are impermeable to liquid (Fig. \ref{fig: cell-schematic}). In these environmental cells, both the liquid and the cell windows introduce undesired distortions to the electron beam as it transmits through the sample lowering the SNR \cite{Pu2020,Ross2015}. Graphene windows are thinner and less dense than commercial silicon nitride windows so introduce less unwanted scattering \cite{Ross2015, Park2021, deJonge2019}. However, even the presence of the liquid alone will lower the SNR compared to the same sample imaged in vacuum \cite{deJonge2018}. To maintain spatial resolution in liquid cell (S)TEM images therefore requires higher electron fluence than achieving the same resolution ex situ in the absence of liquid.

A further challenge associated with liquid cell TEM and STEM is the potential for electron beam induced changes to the liquid cell chemistry. For aqueous solutions, these radiolytic changes to the local liquid environment occur at comparatively low electron fluence \cite{Abellan2014}. The effects include increasing pH and the creation of chemically active free radicals, which may be only indirectly visible in the TEM/STEM images as behavioural artefacts within the system under investigation \cite{Mehdi2021}.  Various methods exist to mitigate radiolytic effects by altering the solution chemistry to inhibit the concentration of radiolytic species \cite{Abellan2014} yet ultimately the coupling of SNR and required image resolution determines the minimum electron dose in the liquid cell \cite{deJonge2018}. Denoising offers the potential for improving the SNR of liquid cell images post-acquisition, enabling reduction in the electron fluence and consequently reducing radiolytic changes in the liquid environment.

Deep-learning techniques have advanced the field of image denoising over the last decade with improved performance, computational efficiency (in application) and by not being reliant on manual parameter tuning \cite{Tian2020}. Deep learning techniques typically require training data, in the form of input and ground-truth/target pairs, to train models. Application to experimental denoising tasks like improving the SNR of TEM images, therefore presents a challenge because often the ground-truth input data (noise-free equivalent of the experimental data) is not known.

There are several examples of supervised deep learning denoising techniques being successfully applied to TEM images \cite{Noise2Atom, Khan2023, Ede2019}. These have been generally found to outperform more traditional approaches such as total-variation (TV) regularisation \cite{Kawahara2022}, non-local low-rank/sparse-representation techniques \cite{Mevenkamp2015, Yankovich_2016} and simple filters \cite{Roels2018}. However, they required extensive training from image simulations or artificially noised images, which can be computationally expensive, and once trained demonstrate limited transferability to different samples. Various examples of the application of denoising to environmental cells can be found in literature \cite{Marchello2020, Reboul2021, Frangakis2021} but the available improvement in SNR was relatively modest despite high computational complexity.

The Noise2Noise family of deep-learning techniques present a powerful alternative solution to the problem of lacking ground-truth (noise-free) equivalents of noisy experimental images but which has not been rigorously investigated for application to denoising of (S)TEM data. Broadly, the Noise2Void approaches use specialised training regimes in order to utilise the noisy experimental images as both their input and ground-truth training data, thereby providing opportunities for image denoising without requiring manual intervention or extensive simulated data to train a model. The Noise2Void technique involves applying these specialised training regimes to a UNet \cite{Noise2Void}, which is a type of convolutional neural network (CNN) that would otherwise typically require supervised training (involving paired/labelled training data) \cite{UNet}.

In this paper we test application of the Noise2Void denoising approach to the exemplar system of atomically resolved STEM imaging of gold adatoms in graphene liquid cells (Fig. \ref{fig: cell-schematic}). The sample is atomically dispersed gold on few-layer graphene, imaged in situ within a graphene liquid cell to visualise the dynamic behaviour at the solid-liquid interface (Fig. \ref{fig: cell-schematic}b). We present denoising of experimental and simulated data, and quantitatively compare denoising performance to simple Gaussian filtering and the more computationally demanding total variation approach. We demonstrate how this denoising approach can be used as an initial step to enhance the effectiveness of feature identification from atomic resolution images, as illustrated in Fig. \ref{fig: cell-schematic}b.

\section{Results}

\subsection{Noise2Void Implementation}

To train a suitable denoising model for a particular set of images, the Noise2Void algorithm assumes that the image-noise is random and pixel-wise uncorrelated \cite{Noise2Void}. For each training datum, the algorithm uses the same noisy image as both input and ground-truth, thereby training the network to replicate the input at the output. Taken alone, this would simply train the model to learn unity, making it useless. The key insight is that a `blind-spot' is introduced to the model, preventing it from using the value of pixel $A$ at the input to predict the value of pixel $A$ at the output. Each pixel value at the output must be predicted using that pixel's neighbours (at the input), but not that pixel itself. Since pixel-wise uncorrelated information cannot, by definition, be predicted with knowledge of a pixel's neighbourhood only, it cannot be reproduced at the model's output. Pixel-wise correlated information in the image can however be reproduced at the output. Therefore, in attempting to reproduce the input at the output, the network reproduces the input image without pixel-wise uncorrelated noise \cite{Noise2Void}. The `blind-spot' effect is not enforced in the network architecture, but is implemented in training. During training, the input image has a grid of pixels masked, with their intensity replaced by some other value.  

The family of Noise2Void inspired algorithms includes the original Noise2Noise, as well as Noise2Self, Noise2Void and N2V2 \cite{Noise2Noise, Noise2Self, Noise2Void, N2V2}. The Noise2Void technique uses a UNet architecture \cite{UNet} and the `blind-spot' masked pixel's intensity is replaced with one of its neighbour's \cite{Noise2Void}. However, the key novelty of the Noise2Void technique itself does not place a strict/specific constraint on the exact network architecture. N2V2 \cite{N2V2} improves on Noise2Void, in part by using a modified UNet, which is the approach we adapt in this work. In N2V2, the masked pixel's intensity is replaced not by a neighbour but with a local average \cite{N2V2}. For both Noise2Void and N2V2, the model's training loss is calculated by determining the model's ability to reconstruct the input image \textit{only at the masked pixels}. Since each masked pixel is modified at the input, their prediction at the output can only be made based on the intensities in their local neighbourhoods, thus implementing the `blind-spot' without needing to add constraints to the network architecture. To successfully apply Noise2Void to denoising of STEM data, several adaptations were made to the Noise2Void techniques described previously in the original Noise2Void and the N2V2 paper \cite{Noise2Void, N2V2}. This included modifications to the input format, the network architecture and the training regime to allow application of the approach to experimental STEM images.

\subsubsection{Modification to support multiple channels}

One of the advantages of the STEM imaging approach is the ability to simultaneously acquire data from different detectors. Information in these images from different detectors is often physically correlated. To illustrate the application of Noise2Void to such data sets we have acquired STEM image pairs, collecting data from the bright field (BF) and annular dark field (ADF) detectors simultaneously (see Fig. \ref{fig: cell-schematic}a). These images are generated from detector measurements acquired simultaneously during a single raster scan and contain physically correlated, yet distinct, information due to the different angular sampling of the transmitted electrons, so should be considered together to gain the best understanding of the material under investigation. Images produced from the ADF detector with high collection solid angle are often referred to as Z-contrast, being dominated by Rutherford scattering with the intensity scaling with the sample's thickness and atomic number \cite{Krivanek2010}. The BF detector collects small angle scattering to produce a phase contrast image or interference pattern where regular crystalline features tend to give strong contrast, although it is not directly interpretable in terms of the crystal structure of the material \cite{Cowley1976}.  Previous implementations of Noise2Void and N2V2 in literature have only applied the technique to monochrome images (i.e. a single colour channel) \cite{Noise2Void, N2V2}. As our experimental data-sets consists of time-series of image-pairs, we have modified the network to operate on both ADF and BF channels in unison by implementing the input frames as dual-channel images. This has the advantage that the network can use BF channel information to enhance its denoising of the ADF channel and vice versa.

\subsubsection{Modification to network architecture}

In this work the UNet architecture was modified compared to the original UNet paper \cite{UNet} and to all previous Noise2Void papers \cite{Noise2Void, N2V2}. The original UNet paper uses an encoder-decoder architecture with a depth of four, max-pooling between layers and 64 feature channels at the output of its first layer \cite{UNet}. The original Noise2Void paper changed the UNet to have a depth of two and 32 feature channels at the output of its first layer (so a much leaner network) \cite{Noise2Void}. The effect of the number of layers in the UNet architecture in dictating the denoising success was tested in the range two to four layers, with four layers found to give the best performance. Therefore for all the results presented in this work, the UNet has a depth of four, with 64 feature channels at the output of the first layer.

The pooling technique was also experimented with, inspired by modifications to pooling reported in the N2V2 paper \cite{N2V2}, with the aim of optimising denoising performance. Replacing the max pooling layers of UNet with average pooling was found to improve denoising performance, likely due to the greater linearity of information-transfer down the network.

A further architecture modification made here is the increase of the receptive field compared to the original Noise2Void implementation \cite{Noise2Void} and the N2V2 \cite{N2V2} architecture. With this change the network has more information on which to make pixel-predictions. Although increasing the number of layers in the network would also achieve this, increasing the number of layers beyond four significantly increases GPU VRAM requirements so the approach becomes harder to practically implement. Increasing the receptive-field in only the first layer was sufficient to achieve improved denoising performance. The size of the convolutional kernel in the first layer was therefore increased to 5x5, with 3x3 kernels elsewhere, whereas other networks use 3x3 kernels in all layers. Our modified Noise2Void architecture is illustrated schematically in Fig. \ref{fig: cell-schematic}a.

\begin{figure}
    \centering
    \includegraphics[width=0.75\textwidth]{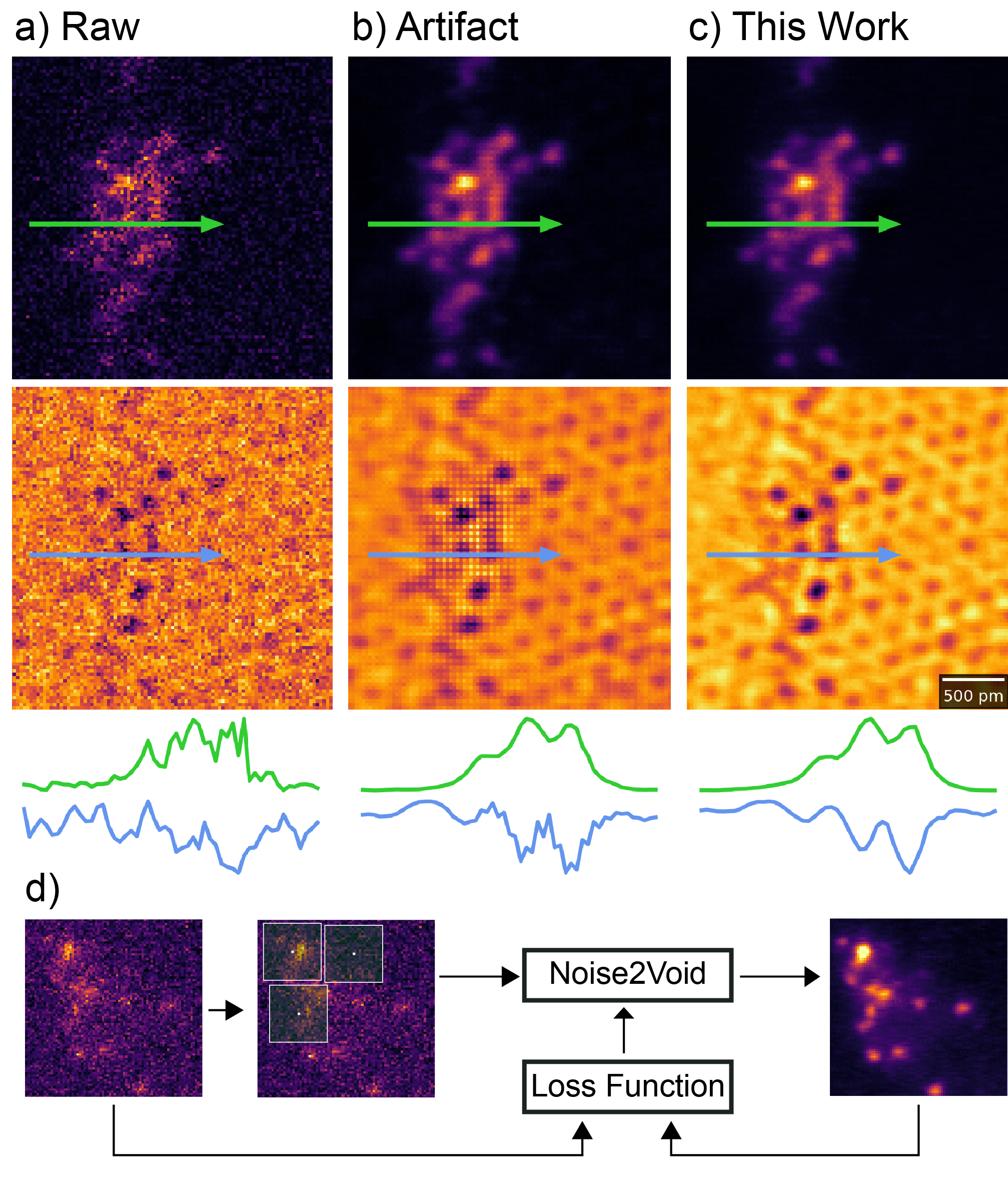}
    \caption{Modified approach to remove chequerboard artifacts found when denoising. (a) Example input dual-channel image. (b) Denoising result showing clear chequerboard artifacts where denoising was performed with an early version of our underlying modified Noise2Void UNet using standard pixel-masking (during training) and transposed convolutions for upsampling. (c) The final Noise2Void UNet architecture, used elsewhere in this work, with chequerboard artifacts significantly diminished. The architectures of (c) and (b) have different upsampling and jittered pixel-masking. An overview of the modified Noise2Void training algorithm can be seen in (d), with the masked pixels jittered/randomly translated by a small amount. Note that, while only annular dark-field images are shown in (d), training is performed on the full dual-channel images.}
    \label{fig: jitter}
\end{figure}

\subsubsection{Modification to training}

Our Noise2Void model was first trained on 8475 dual-channel frames from a full data set of 16750 frames with the training data extracted by sampling every other frame. Initial results with Noise2Void produced grid-like artefacts in the denoised frames (see Fig. \ref{fig: jitter}b), unwanted features which are also reported in the original Noise2Void paper \cite{Noise2Void}. These artefacts are reduced with the addition of a jitter to the positions of the masked pixels (px) in the training images, specifically random displacements of up to 2px translations, and modification to the network's upsampling approach \cite{odena2016deconvolution}. This differs from the N2V2 approach where grid-like artefacts were reduced by removing the outermost skip connection, which in this case could have limited spatial resolution.

Other modifications to the Noise2Void architecture were also tested while optimising the Noise2Void denoising performance, including replacing the convolution blocks with Xception \cite{Xception} style separable convolutions (to improve network efficiency), adding an Atrous Spatial Pyramid Pooling (ASPP) layer \cite{ASPP} at the bottleneck and replacing zevery convolution block with an ASPP block, but all resulted in a poorer denoising result. Removing the outermost skip-connection compared to the original UNet was also tested as reported in N2V2,\cite{N2V2} but this also resulted in worse performance.

\subsection{Denoising performance – experimental data}
\label{ssec: experimental-denoising-performance}

\begin{figure}[H]
    \centering
    \includegraphics[width=\textwidth]{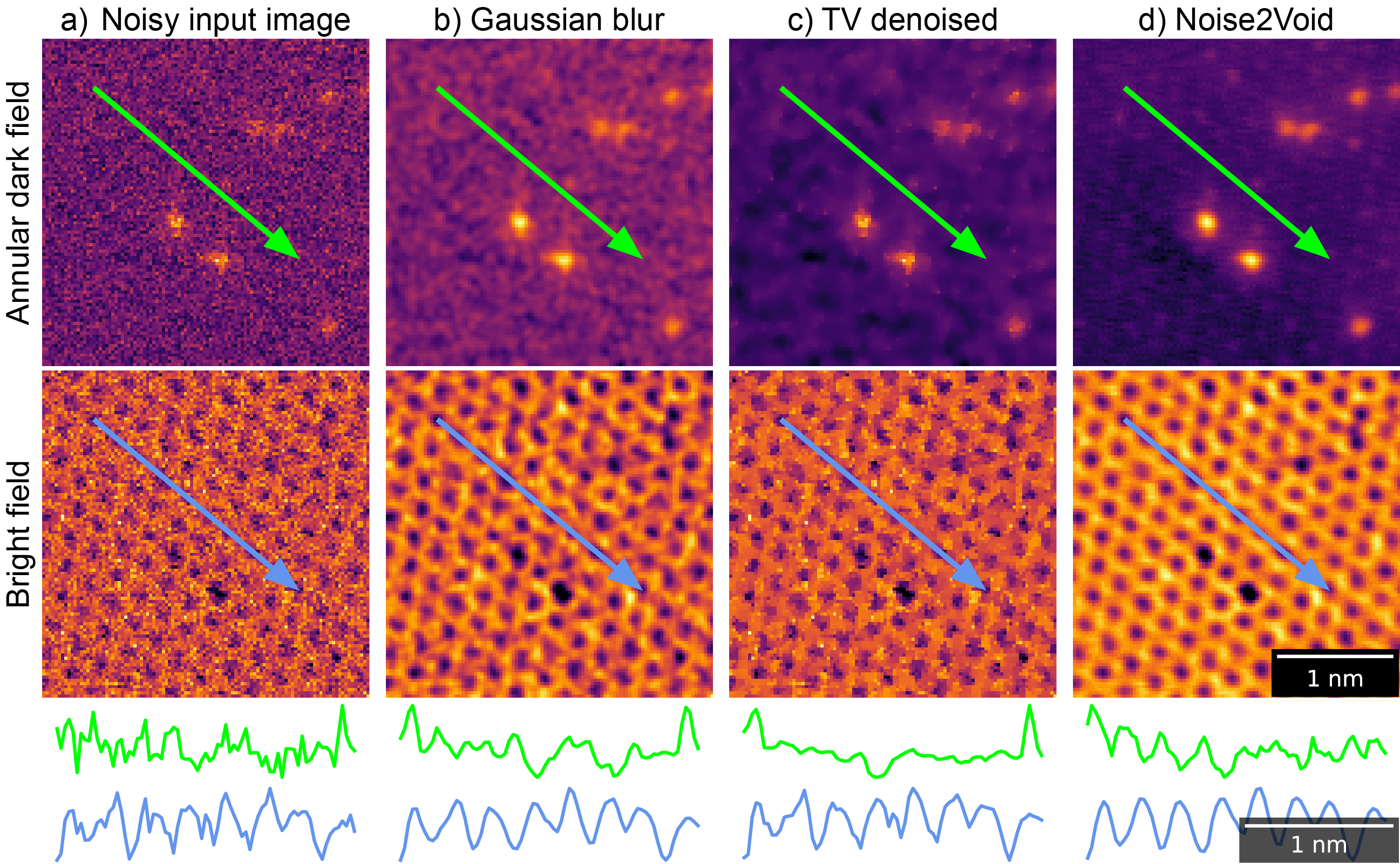}
    \caption{Demonstration of Noise2Void denoising performance on the experimental graphene liquid cell data compared to standard methods. Each frame is a dual-channel image (ADF on upper row, BF on lower row). Intensity line-profiles are presented under the images showing the intensity modulations resulting from the crystal lattice of the graphene windows. The line-profiles are taken along the $[10\bar10]$ direction of the graphene lattice, at the same position for all data (indicated by the green and blue arrows on ADF and BF images respectively), with a width of 1 px. From left to right the images compare the same frame a) from the original experimental data used as input for all the denoisers, b) after Gaussian denoising, c) after denoising by the total variation technique and d) after denoising by our modified Noise2Void approach. The square-root of pixel intensities are displayed for the ADF channel, while the BF images and all line-profiles are plotted on a linear scale. Only Noise2Void denoising recovers the periodic graphene lattice both the BF and the ADF channels.}
    \label{fig:line_profile}
\end{figure}

Fig. \ref{fig:line_profile}a shows a representative frame from the experimental dataset, consisting of a dual-channel ADF-BF image pair with the ADF image shown on the top row and the simultaneously acquired BF image on the bottom row. The aim of the denoising is to improve confidence in identification of the positions of the Au adatoms at the solid liquid interface (bright features in the ADF images and dark features in the BF images) even at low electron fluence. A secondary aim is the identification of the location of the Au adatoms with respect to the underlying graphene lattice, which requires resolving the graphene's atomic lattice. We first consider the latter imaging challenge. Fig. \ref{fig:line_profile}b-d compares the same experimental image pair after Gaussian denoising, TV denoising and after the Noise2Void approach. Unlike our modified dual-channel Noise2Void, Gaussian and TV denoising are not dual-channel techniques, and so are applied to each image channel separately. To more clearly highlight the different SNRs, intensity profiles are extracted from identical locations on all images demonstrating the intensity modulation resulting from the graphene lattice sampled along the $[10\bar10]$ direction as shown in Fig. \ref{fig:line_profile}a-d. Examination of Fig. \ref{fig:line_profile}d shows that the UNet Noise2Void approach has revealed the underlying graphene lattice in both the BF and the ADF channel, while the periodicity is not clear in the ADF images subjected to either Gaussian blurring or TV denoising (Figs. \ref{fig:line_profile}b-c).

\begin{figure}[H]
    \centering
    \includegraphics[width=\textwidth]{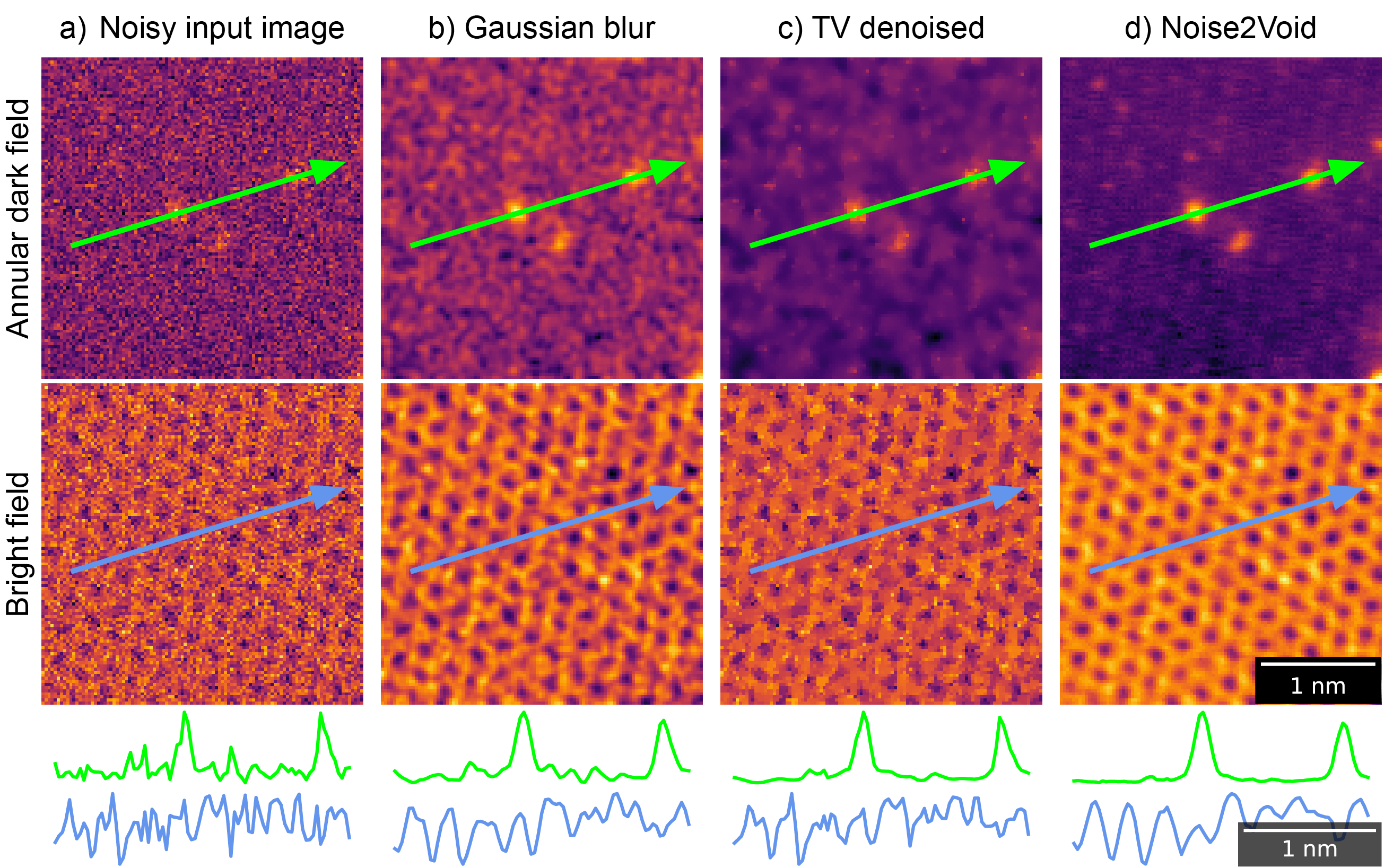}
    \caption{Demonstration of Noise2Void denoising performance on the experimental graphene liquid cell data (ADF and BF image pairs) and comparison to standard denoising methods. Intensity line-profiles are presented under the images showing the intensity modulations. The line-profiles are taken through two gold adatoms and are at the same position for all data, (indicated by the green and blue arrows on ADF and BF images respectively) with a width of 1 px. From left to right the images compare the same frame a) from the original experimental data used as input for all the denoisers, b) after Gaussian denoising, c) after denoising by the total variation technique and d) after denoising by our modified Noise2Void approach. The square-root of pixel intensities are displayed for the ADF channel, though the BF images and all line-profiles below are plotted linearly. The ADF has a much flatter background with the Noise2Void denoising, compared to the input, Gaussian blur and TV denoising, making it easier to recover the Au adatom positions.}
    \label{fig: line-profile-cropped}
\end{figure}

Once the graphene lattice is resolved it is necessary to identify the number and position of Au adatoms on the graphene surface. Fig. \ref{fig: line-profile-cropped} provides similar data to Fig. \ref{fig:line_profile} but with the line profile sectioning through a Au adatom on the graphene inside the liquid cell.

The TV and Noise2Void denoisers produce flatter `background' (Au-free graphene region) intensities in the ADF image than Gaussian blurring, which has the benefit of increasing the number of Au adatoms that are correctly identified (as shown in Fig. \ref{fig: cell-schematic}b).

\clearpage

A rigorous quantification of SNR in TEM images requires the underlying ‘ground-truth’ image signal be known, to accurately measure the noise. This is achievable by comparison to image simulations as demonstrated in section \ref{ssec: simulated-denoising-performance}. Nonetheless, useful insights can still be gained from consideration of experimental SNR values.

We first consider Noise2Void denoising of the experimental ADF images. The background ‘noise’ includes a ‘real’ periodic signal from the graphene lattice as well as unwanted intensity variations resulting from both the presence of the liquid and background noise from the imaging system. The Z-contrast dependence of the ADF signal means the graphene lattice is only a weak modulation compared to the intensity of Au adatoms, so when seeking to identify only the Au adatom peak locations it can be considered small compared to the unwanted signals we wish to remove. In practice, Noise2Void denoising can reveal the periodic modulations of the graphene lattice in the ADF signal (as shown in Fig. \ref{fig:line_profile}d). However, we ignore this in the following analysis, recognising this lattice signal will result in underestimation of the SNR improvement when gold-free regions are considered as the noise reference.

\clearpage

The signal to noise ratio (SNR) for the Au adatoms measured from the ADF intensity line profile in Fig. \ref{fig: line-profile-cropped} gives a value of 2 for the raw input image. This is improved to 3.3, 7.2 and 12 for Gaussian blurring, TV denoising and Noise2Void denoising, respectively (see methods for details of SNR quantification). Arguably more challenging than identification of the positions of well isolated individual adatoms, is the ability to resolve separate adatoms that are in close proximity. This competes with achieving a high SNR for single adatoms when using Gaussian blur so may be a limiting factor for conventional denoising approaches. Supplementary information (SI) Fig. S\ref{fig: si-line-profile} compares intensity line profiles where two Au adatoms are separated by 0.34 nm ($\sim7.4$ px). Quantification of the drop in the intensity of the ADF image between the adatoms, relative to the height of the least intense adatom peak reveals a drop of 57\% for the original raw input image, compared to 28\%, 14\% and 29\% for the Gaussian blur, TV and Noise2Void respectively. Together, the data in Figs. \ref{fig: line-profile-cropped} \& S\ref{fig: si-line-profile} suggests that Noise2Void is the better denoiser for the ADF channel because it improves the SNR significantly from 2 to 12, while also retaining the ability to separately distinguish nearby features, which can be suppressed by other denoising methods.

Compared to the ADF image-channel, the denoising performance for the BF channel in Fig. \ref{fig:line_profile} is harder to assess because the feature of interest (the graphene lattice) covers the whole image and there is no region that can reasonably be approximated as purely background. Considering Figs. \ref{fig: line-profile-cropped} \& S\ref{fig: si-line-profile}, in all denoisers, but especially TV and Noise2Void, higher-frequency components have been diminished yet it is not clear if this significantly improves the image for the extraction of the crystallographic lattice in later analyses.

\begin{figure*}
    \centering
    \includegraphics[width=\textwidth]{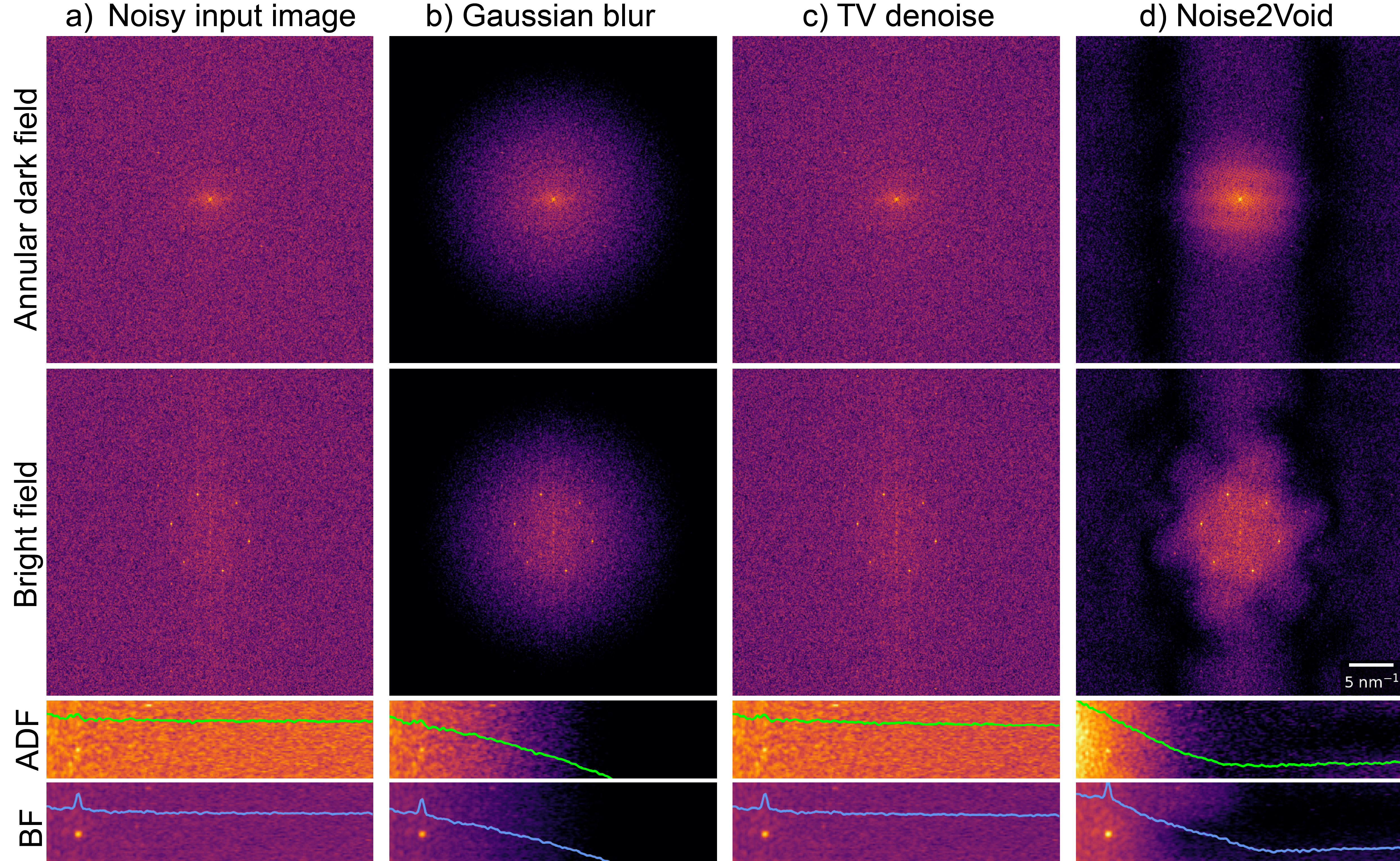}
    \caption{Fourier space demonstration of Noise2Void denoising performance on the experimental graphene liquid cell data and comparison to standard denoising methods. The magnitudes of 2D fast Fourier transforms (FFTs) of the ADF and BF image channels (first row and second row, respectively) are shown with  intensities plotted on a logarithmic scale. The third and fourth rows show a polar transform of the 2D FFTs, for the ADF and BF images respectively, six-way folded in the azimuthal direction with the azimuthally integrated intensity overlaid. For each row, intensities are displayed equally for ease of comparison. From left to right the columns compare the same frame a) from the original experimental data used as input for all the denoisers, b) after Gaussian denoising, c) after denoising by the total variation technique and d) after denoising by our modified Noise2Void approach.}
    \label{fig:mag_spectrum}
\end{figure*}

The transfer of periodic crystallographic information after denoising is more easily assessed in Fourier space as shown in Fig. \ref{fig:mag_spectrum}, where the upper row is the magnitude of the fast Fourier transform (FFT) of the ADF image and the second row is the FFT-magnitude of the corresponding BF image.  The input frame is shown in Fig. \ref{fig:mag_spectrum}a, while Fig. \ref{fig:mag_spectrum}b-d shows each denoiser's output frame. The Gaussian denoised FFT-magnitude shows that only high spatial frequency noise has been removed from both the ADF and the BF images, as expected with this simple approach. Encouragingly, the Noise2Void approach has both suppressed the high frequency noise and also enhanced the transfer of low spatial frequencies relative to the high spatial frequencies in both ADF and BF images. The graphene lattice appears in the FFT-magnitudes as sets of hexagonal lattice spots. These are visible in the FFT-magnitudes of the BF input image and in all the denoised BF images. However, in agreement with the real space analysis in Fig. \ref{fig:line_profile}, only denoising with Noise2Void reveals all six of the first order graphene spots (corresponding to the $\{1\bar100\}$ lattice spacings) for the FFT of the ADF image.

The third and fourth rows of Fig. \ref{fig:mag_spectrum} show six-fold reduced polar plots of the FFT-magnitudes, where these have been warped by a polar transformation (such that the horizontal axis now corresponds to the radial direction in the above FFT-magnitudes) and summed periodically over $60^{\circ}$ (azimuthal) sections/wedges. Line profiles have then been overlain, where they display the totally azimuthally integrated intensity as a function of radius (the radial profile). In these radial profiles each set of six-fold symmetric hexagonal graphene lattice spots reduces to a single spot. While the lowest order graphene lattice ($1\bar100$ spot) is clearly visible in the BF radial profile for all data presented, careful inspection will also reveal these spots in the ADF radial profiles. The higher order ($11\bar20$ type) spots can also be seen in both channels which, for the BF channel have SNR values of $9.0$, $8.2$, $10.7$ and $16.0$ for the input, Gaussian blur denoised, TV denoised and Noise2Void denoised images respectively.

The Noise2Void ADF FFT-magnitude shows an unusual hexagonal symmetry in the information transfer, approximately corresponding to the orientation of the graphene lattice.  It appears that Noise2Void has selectively suppressed regions in the FFT more distant from the lattice-points of the graphene lattice, in a manner akin to a process that is often performed manually by selective FFT filtering to enhance the visibility of a crystal lattice. Nonetheless, no component of the underlying UNet architecture is six-fold rotationally symmetric, meaning the Noise2Void model has learned this symmetry during training. While denoising of a perfect lattice can be effectively achieved with manual FFT filtering, this utilises prior knowledge of the sample's structure to choose how and where to modify the FFT (i.e. a physics-based model). In contrast, Noise2Void has no initial knowledge of the sample and must therefore have learned the intrinsic six-fold symmetry by identifying correlations within the videos (i.e. a data-driven model). The hexagonal symmetry introduced by Noise2Void denoising accommodates the various orientations of the crystal lattice across the experimental dataset. The resemblance of the FFTs of Noise2Void's denoised videos to what one might expect from an expert's FFT filtering can be seen as endorsement of the ability of Noise2Void to identify and patterns in the training data and exploit that correlation to enhance denoising performance. Expert FFT filtering is often laborious (requiring adaptation whenever the sample's orientation changes) whereas Noise2Void's results in similar success but without manual intervention.

\subsection{Denoising Performance - Simulated Data}
\label{ssec: simulated-denoising-performance}

\begin{figure}[h]
    \centering
    \includegraphics[width=\textwidth]{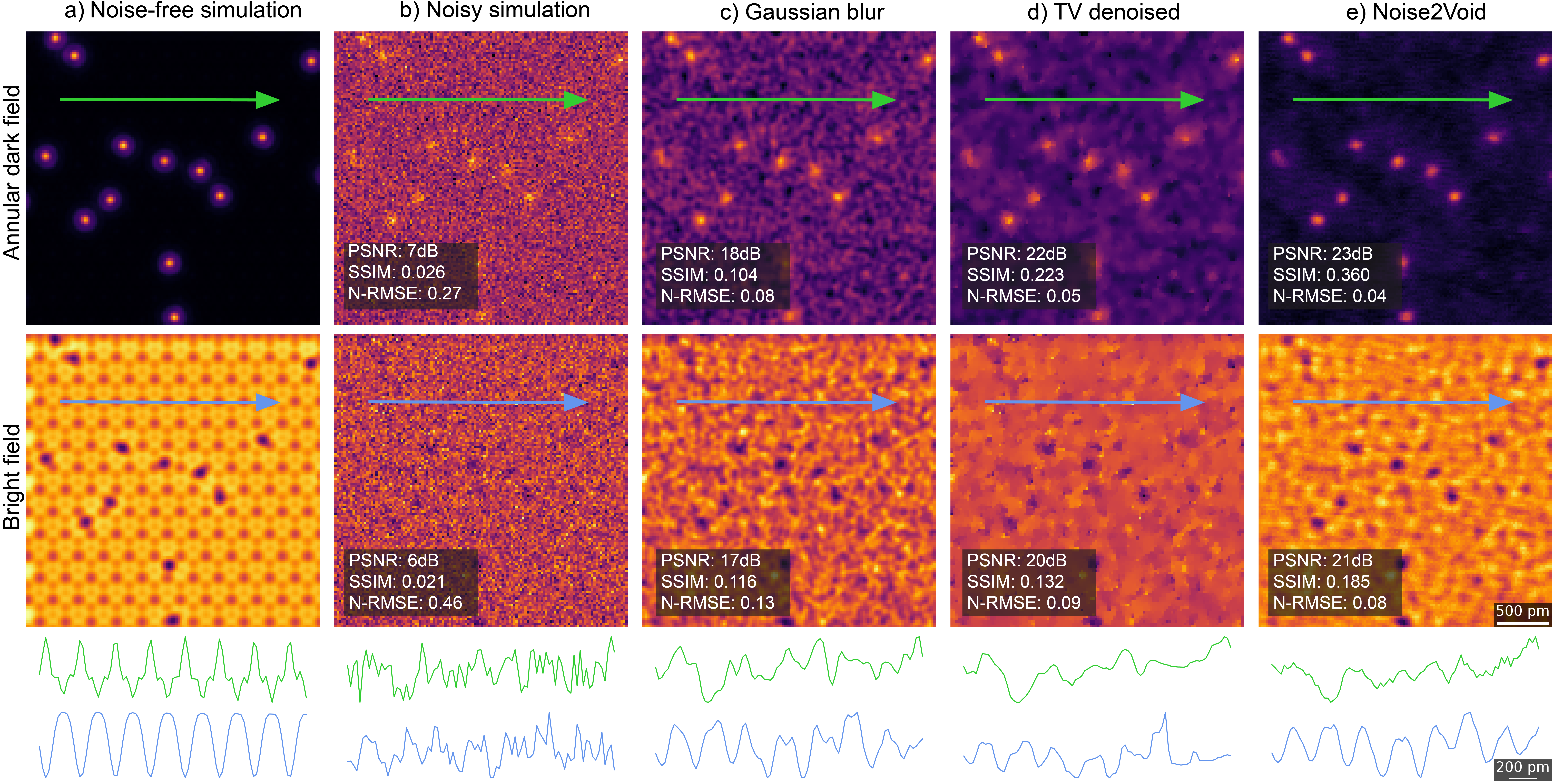}
    \caption{Demonstration of the denoising performance of the three denoising techniques applied to a simulated liquid-cell frame. Frames are dual-channel annular dark-field (upper row) and bright-field (second row) with input PSNR of 7 dB and 6 dB, respectively. From left to right the columns compare a) the simulated frame without noise, b) simulated frame with noise added used as input for all the denoisers, c) after Gaussian denoising, d) after denoising by the total variation technique and e) after denoising by our modified Noise2Void approach. Below the images are plotted the intensity line-profiles extracted with a width of 1px.
    }
    \label{fig: sim_denoised_comparison}
\end{figure}

\begin{figure}
    \centering
    \includegraphics[width=0.5\linewidth]{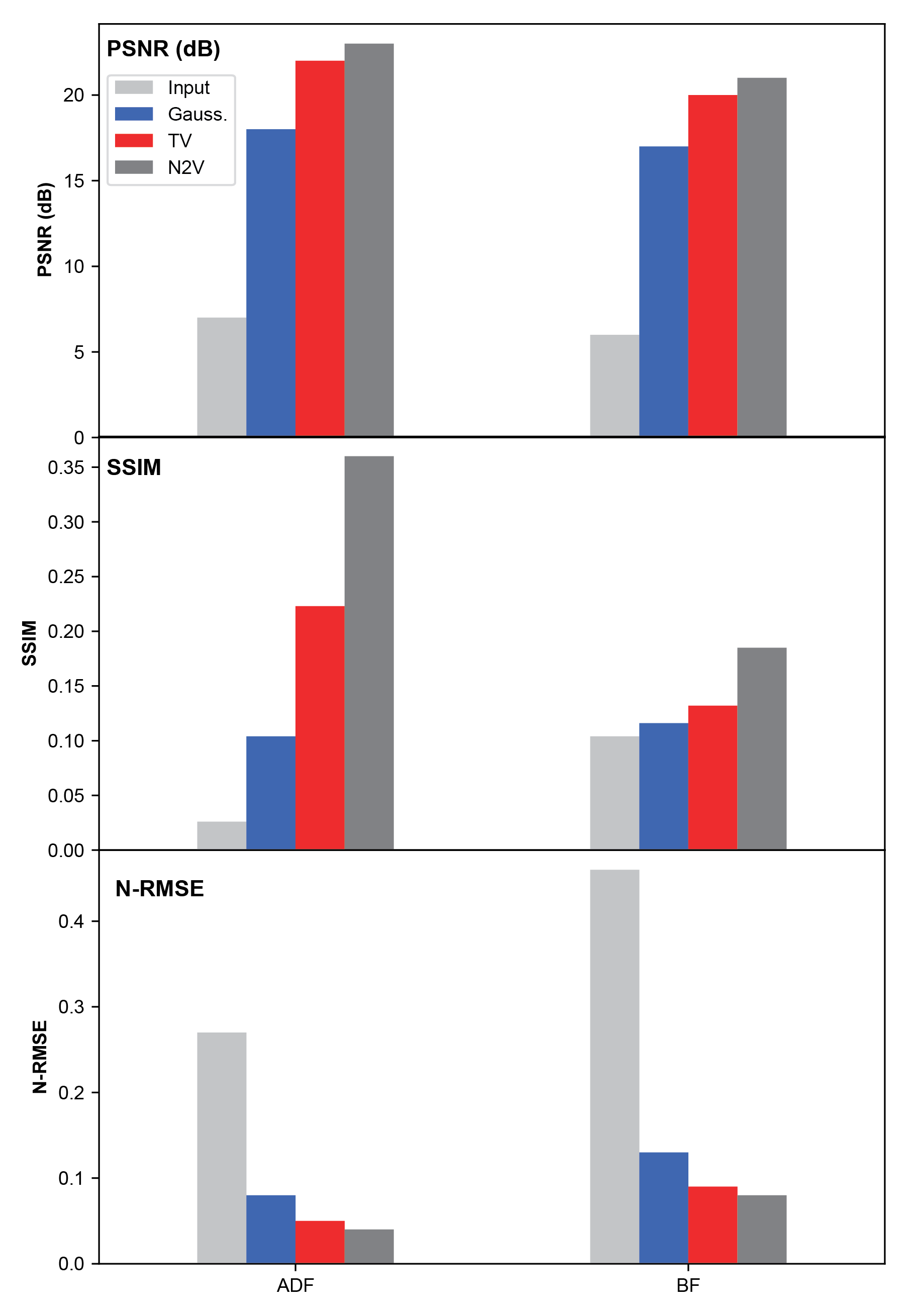}
    \caption{Evaluation of the denoising performances of a simple Gaussian blur, TV denoising and Noise2Void applied to a simulated liquid-cell frame. Graphs show how denoising metrics separately for ADF and BF channels (corresponding to the images shown in Fig. 6). These metrics compare the similarity of each denoiser’s output (and the noisy input) to the noise-free simulation. For PSNR and SSIM, larger values are better while for N-RMSE smaller values are better.}
    \label{fig: denoising-performance-barcharts}
\end{figure}

Despite the apparent success of the Noise2Void method presented in Figs. \ref{fig:line_profile}, \ref{fig: line-profile-cropped}, \ref{fig:mag_spectrum} and \ref{fig: si-line-profile}, the lack of ground truth data for experimental images limits quantitative comparison of the different approaches. We therefore apply our Noise2Void approach to denoising of simulated STEM image pairs that closely resemble the experimental data and compare the results to noise-free simulations (see methods for further information). Three metrics are used to allow quantitative comparison; peak signal-to-noise ratio (PSNR), structural similarity (SSIM), and normalised root-mean-square error (N-RMSE). Peak signal-to-noise ratio (PSNR) and structural similarity (SSIM) indicate the similarity of the image to the ground-truth (higher is better), while normalised root-mean-square error (N-RMSE) indicate the difference from the ground-truth (lower is better). While PSNR is the most widely used metric, SSIM is more closely correlated with human perception and N-RMSE is comparatively simpler to understand \cite{Wang2009}. The implementation for each can be found in scikit-image's `metrics' module \cite{scikit-image}. The denoising results for simulated data with an input PSNR value of 7 dB and 6 dB for the ADF and BF channels respectively, are presented in Fig. \ref{fig: sim_denoised_comparison} where our Noise2Void approach is again compared to Gaussian and TV denoising. The PSNR, SSIM and N-RMSE values are summarised in Fig. \ref{fig: denoising-performance-barcharts}. This quantitative comparison shows that the Noise2Void approach gives the best denoising performance for both BF and ADF images, even for very noisy simulated input data, with all metrics significantly outperforming Gaussian blurring and the TV approach for our simulated data. Similar behaviour is seen when considering the SSIM and N-RMSE values with the noisy input data showing an order of magnitude improvement in all metrics for the ADF channel. The SSIM and N-RMSE improvements for the BF channels are also similarly significant, corresponding to $9$ and $6$ fold improvement factors respectively.

\clearpage

\subsection{Denoising at higher PSNR}

\begin{figure}
    \centering
    \includegraphics[width=\textwidth]{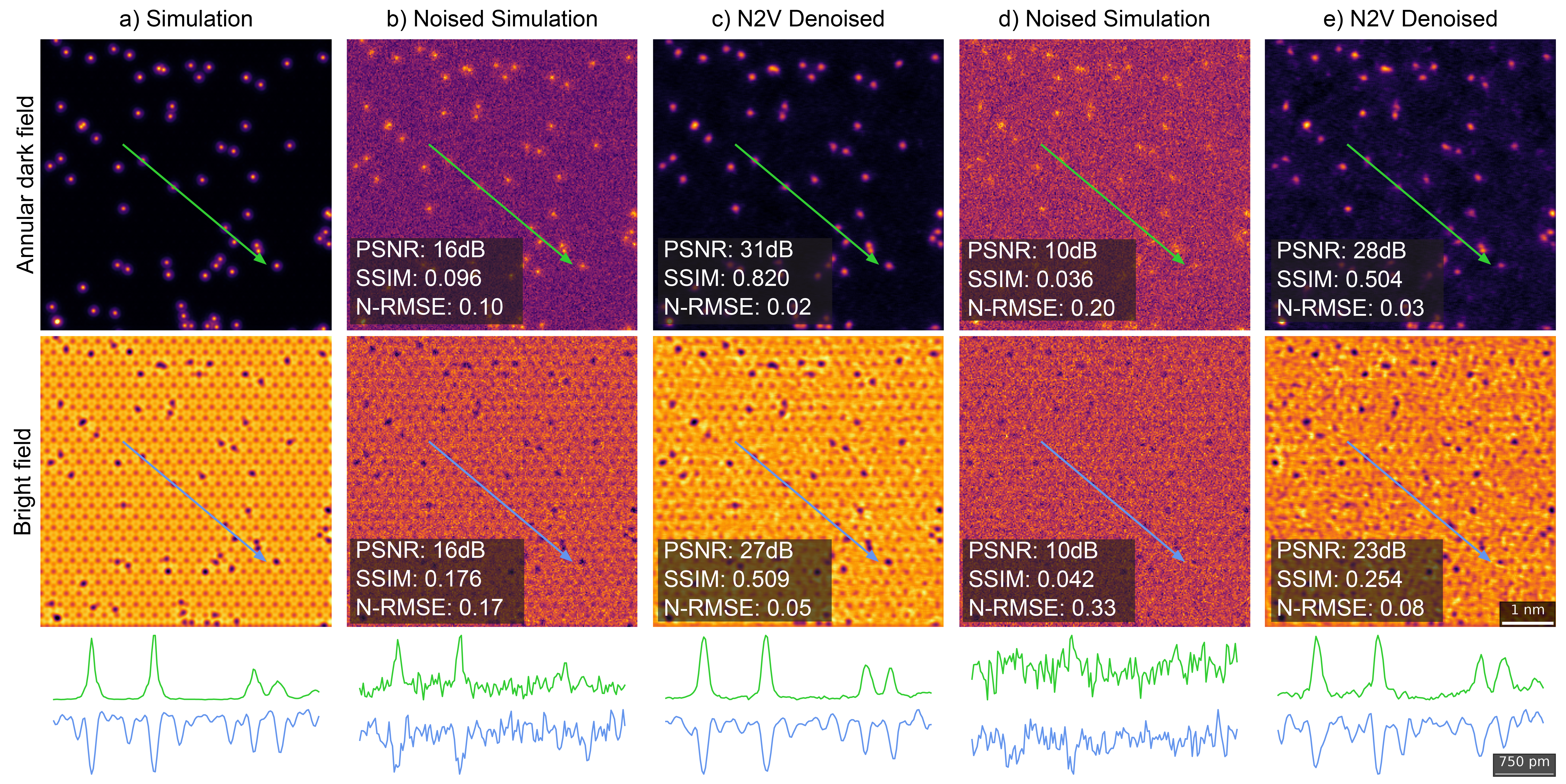}
    \caption{Demonstration of the denoising performance for Noise2Void for simulated liquid cell data with different PSNR values. Frames are dual-channel annular dark-field (upper row) and bright-field (second row). From left to right the columns compare a) the simulated frame without noise, b) simulated frame with noise added to give PSNRs of $16$ dB in the ADF and BF image channels, c) after denoising of the frame in b) with our modified Noise2Void approach, d) simulated frame with noise added to give PSNRs of $10$ dB in the ADF and BF image channels, respectively, e) after denoising of the frame in d) with our modified Noise2Void approach. Below the images are plotted the intensity line-profiles extracted through four Au adatoms with a width of 1px.}
    \label{fig: denoise-series}
\end{figure}

\begin{figure}[H]
    \centering
    \includegraphics[width=0.6\linewidth]{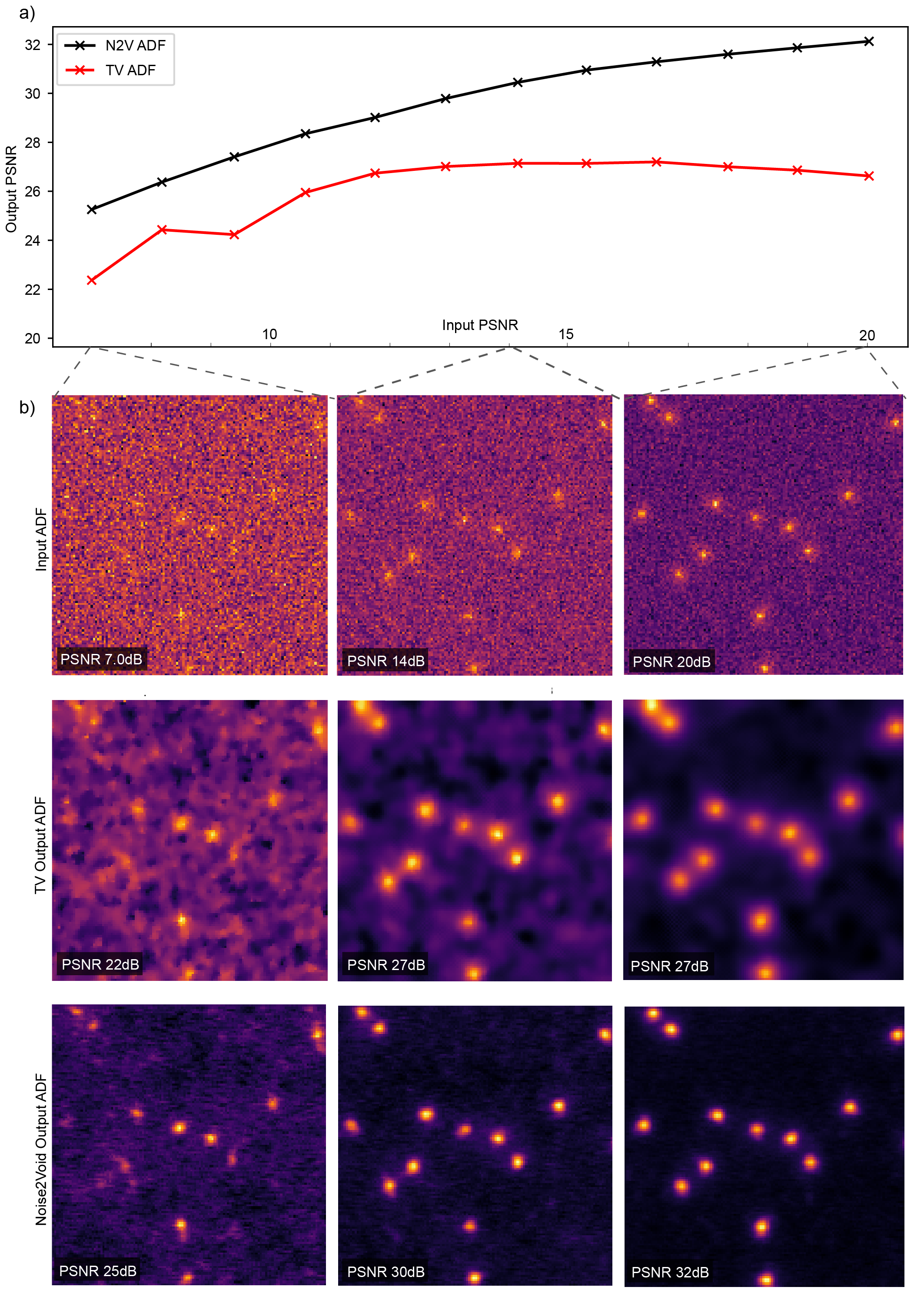}
    \caption{Noise2Void and TV denoising performance as a function of input noise, for the ADF channel. The same noise-free simulated image is used for all noisy inputs, with varying levels of AGWN added. Corresponding input and output images can be seen in SI Figs. \ref{fig: denoiser-performance-profile-inputs}, \ref{fig: denoiser-performance-profile-n2v-outputs}, \ref{fig: denoiser-performance-profile-tv-outputs}.}
    \label{fig: denoiser-performance-profile}
\end{figure}

The experimental input images used in section \ref{ssec: experimental-denoising-performance} have relatively high SNR ($\sim2$ in the ADF images, qualitatively corresponding to a PSNR of $\sim16$ in the simulated ADF images, an order of magnitude higher than the input simulations in Fig. \ref{fig: sim_denoised_comparison}). This high SNR has the advantage that it provides guidance for feature identification from the raw data, yet it is highly desirable to collect raw experimental data with much poorer SNR to minimise the electron dose applied to the sample. To enable optimisation of imaging conditions and infer the minimum electron fluence for data acquisition, we now consider denoising of data for a range of different input PSNR values. The PSNR scale (when measured in dB) is logarithmic so our trained Noise2Void algorithm was applied to simulations with input PSNRs in the range from $7$ dB to $20$ dB. Fig. \ref{fig: denoise-series} demonstrates denoising of input simulated ADF images with PSNR values of $16.0$ and $9.9$ dB, showing improvements to $30.9$ and $27.5$ dB, respectively, in the denoised ADF outputs. A quantitative comparison of the denoising performance for PSNR values from $7$dB to $20$dB and a comparison to TV denoising is shown in Fig. \ref{fig: denoiser-performance-profile} (and SI Fig. \ref{fig: si-denoiser-performance-profile}). Our Noise2Void approach is shown to outperform TV denoising in the ADF channel for all input PSNRs, with the most effective denoising achieved for the ADF channel over the BF. Furthermore, the output PSNR in the denoised image remains exceptionally high across all input PSNRs, giving an improvement ratio of the logarithmic PSNR (PSNR output dB / PSNR input dB) of $\sim3$ at the lowest input PSNR, compared to an improvement ratio of $\sim1.6$ for the highest input PSNR, both for the ADF channel.

\clearpage

\subsection{Comparison of Denoising Speed}

The speed of a particular denoising method is important if it is to be applied for real time denoising as well as for efficient data processing post-acquisition. A significant speed milestone for TEM denoising is therefore matching the rate at which experimental data is acquired, allowing the method to be used to direct or assist with the experimental data collection.

The precise speed of a denoiser will depend on several factors, including disk read-write speed, processor performance, hardware utilisation etc. Nevertheless, it is informative to compare the relative speed of each denoising technique applied to the same video series. The time required for Gaussian blurring, TV and our Noise2Void technique was compared for the denoising of a single video (containing about $\sim$100 frames) on the same hardware. The mean time-per-frame of the modified Noise2Void approach was found to be 22ms (45 frames-per-second), only twice the 11ms required by simple Gaussian blurring, and an order of magnitude less than the 300ms required by TV. The test was repeated 3 times to assess the variability of the time required for each denoising approach and the results were found to vary by approximately 10\%. All three approaches are faster than the relatively slow data acquisition rates used for these liquid cell STEM experiments ($\sim$1.9 frames-per-second). Nevertheless faster acquisition rates of $10-20$ frames-per-second ($50-100$ ms per acquisition) are routinely used in TEM and STEM imaging, for example during instrument alignment and for analysing electron-beam sensitive samples. Noise2Void therefore has the advantage compared to TV denoising of being fully compatible with real-time STEM imaging, suggesting it could provide a valuable tool to assist with alignments, especially when imaging electron beam sensitive samples or where low-dose conditions are required.

\subsection{Transfer Training}
\label{ssec: transfer-training}

\begin{figure*}
    \centering
    \includegraphics[width=\textwidth]{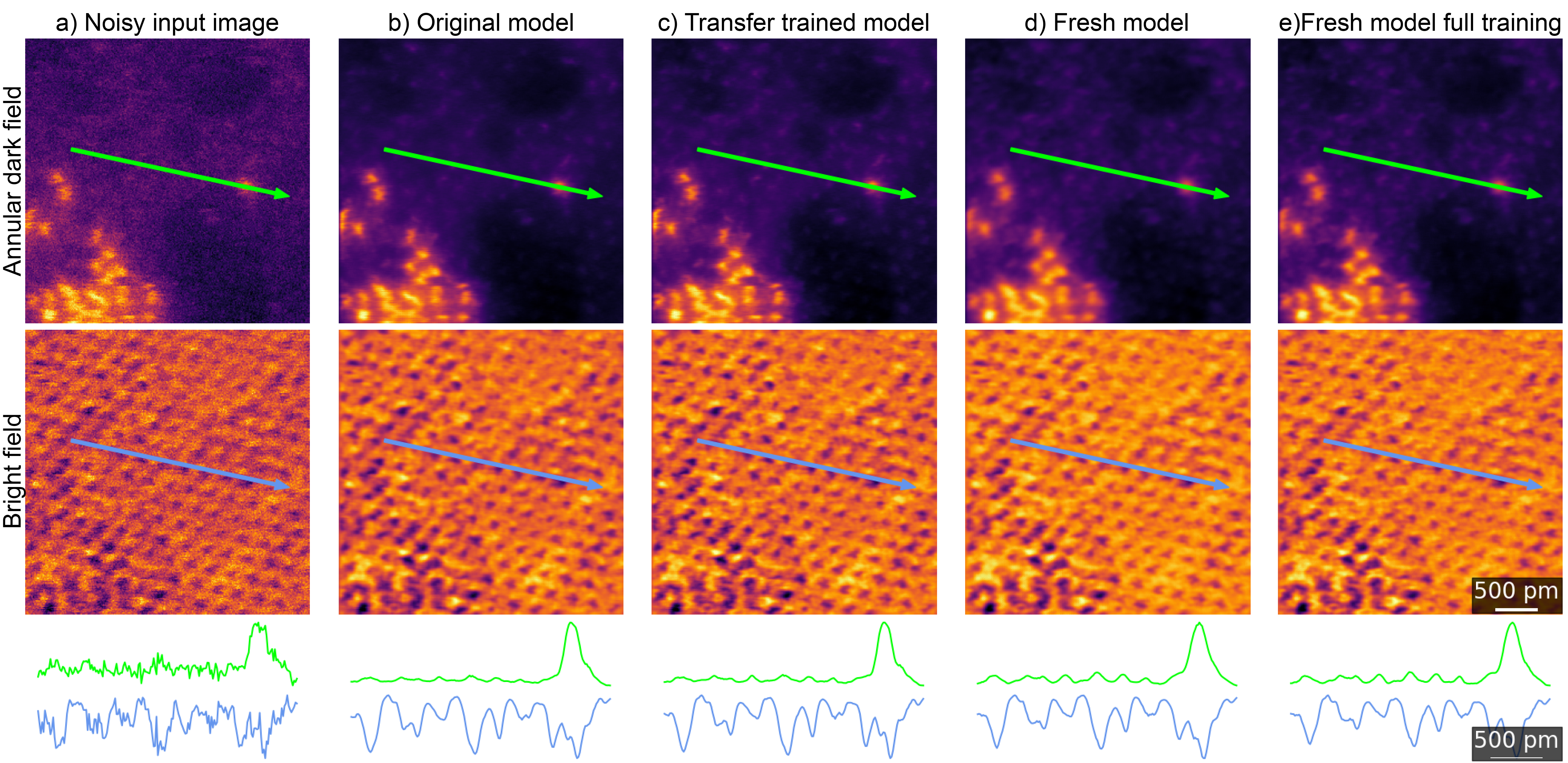}
    \caption{Demonstration of the effectiveness of denoising a new set of experimental data. Frames are dual-channel annular dark-field (upper row) and bright-field (second row). Below the images are plotted the intensity line-profiles (width of 1px), extracted at identical locations along the armchair crystallographic direction of the underlying graphene lattice and bisecting a Au adatom. From left to right the columns display a) the noisy input frame, b) the output using the Noise2Void model trained on previous data (used in Figs. \ref{fig: jitter}-\ref{fig: denoise-series}), c) the output after transfer-training the  original model on this new experimental data, d, e) the output using our Noise2Void network architecture but trained on this new data, using half the data as input (d) and using all of it (e).}
    \label{fig: transfer_comparison}
\end{figure*}

Training of the Noise2Void model is the slowest and most computationally intensive part of the Noise2Void denoising. We find that training requires a high-performance workstation while the denoising can be done on a standard laptop. The time required to train a Noise2Void model is highly dependent on the compute resources available and amount of training data. Approximately an hour and a half was required to train each Noise2Void model, with more details available in the supplementary information. The ability of a trained model to be used for a new experimental data set is therefore important for the wider usability and applicability of this denoising approach. Fig. \ref{fig: transfer_comparison} shows the success of the trained Noise2Void approach when applied to a new set of experimental data. The new experimental data set was 1230 experimental frames which were again dual channel images, 512 x 512 px in size, acquired on a similar but not identical sample. The data was obtained by a different microscopist, and during a different experimental session so instrumental parameters will have changed including lens aberrations, which dictate information transfer to the image. Fig. \ref{fig: transfer_comparison}a shows an example frame from the new experimental data. Fig. \ref{fig: transfer_comparison}b shows the result from our modified Noise2Void model (presented in Figs. \ref{fig:line_profile} \& \ref{fig:mag_spectrum} and trained on previous experimental data) when applied to denoise the new data. Fig. \ref{fig: transfer_comparison}c compares this to where the original model was used as a starting point, but then further trained using a subset of the new experimental data (so overall the model had been trained on both old and new datasets). We refer to this as a `transfer trained' model. Finally, we consider the model performance relative to where completely fresh models were trained only on the complete new data set (Fig. \ref{fig: transfer_comparison}d) and only on half the new data set (Fig. \ref{fig: transfer_comparison}e). Examination of the images and line-profiles demonstrates that all approaches denoise the data effectively with the ADF SNR ratio improved for all the denoised data. The model trained on fresh data gives a slightly improved transfer of the graphene lattice in the ADF channel after denoising but the improvement is small demonstrating that retraining is not required as part of a workflow for similar samples. 

\subsection{Comparison with mono-channel networks}
\label{ssec: mono-channel}

A significant difference of this work over earlier implementations of Noise2Void is the application to dual channel images. It is therefore interesting to consider whether this dual channel UNet provides improved performance compared to two single-channel UNets for denoising dual channel data. To test this, a Noise2Void denoiser was created using two separate mono-channel UNets each trained on a single channel (ADF or BF images). Both networks are identical, except for the number of channels at the input and output, and each mono-channel model was trained for the same number of training epochs as the dual-channel model used elsewhere in this work. Comparison of the training losses for both the dual and the monochannel networks (SI Fig. \ref{fig: training-losses}) shows that the dual channel and ADF monochannel plateaux after 32 epochs but the BF mono-channel is continuing to decrease, suggesting further training is required. 

While the ADF mono-channel showed similar results to the dual-channel Noise2Void denoiser, image artifacts were present in the output of the BF mono-channel denoiser, which is to be expected when the model's training has not converged. While further training of the BF mono-channel network may remove these artifacts, it's presence suggests that the mono-channel denoising technique requires more training compared to the dual-channel model. This would require more compute resources and is therefore disadvantageous.

The speed of the dual-channel denoiser was also compared to the mono-channel, and found to be around a factor of two faster with frames taking on average 43 ms per frame for the mono-channel compared to 23 ms per frame for the dual-channel denoiser (both times are averages over 100 frames). This is unsurprising since for the mono-channel case there are two UNets rather than one. The combination of longer training and slower denoising speed demonstrate that the dual-channel UNet network architecture is preferred for denoising this dual-channel video dataset compared to two mono-channel UNets. Given the increasing capabilities of STEM instruments to collect large multiple-channel data sets, the demonstrated denoising approach for efficient data analysis is likely to be increasingly important.

\section{Conclusions}

In this study, we have demonstrated the successful application of a modified Noise2Void architecture for denoising atomic resolution STEM images. Neither the training nor the application of the method requires manual intervention, with no prior sample knowledge or image simulations required. The method achieved a factor of 3 improvement in PSNR and an order of magnitude reduction in N-RMSE values, as measured by application to image simulation. The success when applied to experimental data is demonstrated by the increased SNR for visibility of both isolated metal adatoms and the graphene lattice, outperforming Gaussian blurring and the TV approach. The Noise2Void methodology is found to have a mean processing time-per-frame of just 23ms (or 45fps) making it fully compatible with real time denoising of experimental STEM images, even when performing live search and alignment procedures. Denoising is highly desirable to improve the SNR of experimental data, providing opportunities for greater information recovery, for lowering the electron flux or for taking experimental data sets at higher frame rates. These new imaging capabilities can be used to reduce the potential for electron beam induced changes to the system or for investigating atomic scale dynamic behaviour with improved temporal resolution. We demonstrate the potential of the technique for atomic resolution imaging of Au adatoms on graphene surrounded by liquid, but we expect that the presence of the liquid does not affect the denoising and that the Noise2Void approach we describe is broadly applicable to denoising of any atomic resolution STEM images.

\section{Methods}

\subsection{Sample Preparation}

The cells were constructed using layer by layer transfer of 2D crystals on to a SiN support grid as described previously \cite{Kelly2018, Clark2022}. The cell was filled with a solution of $\mathrm{HAuCl}_4$ salt dissolved in organic solvent to generate atomically dispersed species of Au and Au clusters on a graphene surface surrounded by liquid. The upper and lower graphene windows are few layer graphene (3-4 layers or $\sim$1nm thick) and the liquid thickness is controlled by the thickness of the boron nitride spacer crystal to be 30-50 nm (Fig. \ref{fig: cell-schematic}). The samples were imaged using a double corrected GrandARM JEOL ARM300CF STEM with a convergence angle of 32 mrad, an accelerating voltage of 80kV. Simultaneous bright field and annular dark field STEM images were acquired with acceptance angles of up to 18.7 mrad and between 46.8mrad and 169.3 mrad, respectively. Frame sizes were $512\times512$ pixels. In total 16950 frames were acquired as video sequences with $\sim$100 frames per video.

\subsection{Multislice Simulations}

Multislice image simulations were performed using the abTEM software \cite{abTEM}. Atomic coordinates were for a 4-layer sheet of Bernal stacked graphene, with gold atoms randomly deposited on the upper surface. Simulations parameters were matched to the JEOL GrandARM experimental data.

White noise (Gaussian) noise was added to the simulated images in Fig. \ref{fig: sim_denoised_comparison} as additive Gaussian white noise (AGWN). PSNRs were simulated in the range of $\sim7-20$dB, with values of of 16.4  and 16.2 for the ADF and BF channels respectively, found to qualitatively provide the best match to the SNR of the experimental images.

\subsection{Denoising}

All denoising processing was performed on a Nvidia A5000 workstation. Gaussian blurring was performed frame-wise and channel-independently with a standard deviation of the Gaussian kernel of $\sigma = 1\mathrm{px}$ throughout the dataset. Kernels of $\sigma = 2\mathrm{px}$ and $\sigma = 0.5\mathrm{px}$ were also tested but found to give poorer performance so are not included in the results.

Total variation denoising was performed using the Chambolle algorithm \cite{Chambolle2004} as implemented by \textit{scikit-image} v0.20.0. This implementation takes as a parameter $\frac{1}{\lambda}$ which was chosen as $1.0$ for the ADF channel and $0.75$ for the BF channel. These values were chosen as optimal after performing denoising calculations for $\frac{1}{\lambda}$ = $0.1$ - $2.0$ in increments of $0.1$.

For the Noise2Void denoising the UNet has a depth of four, with 64 feature channels at the output of the first layer. Average pooling is used and the size of the convolutional kernel in the first layer is increased to 5x5 kernels with 3x3 kernels used elsewhere. Our modified Noise2Void model was initially trained on 8475 frames from a full data set of 16750 frames with the training data extracted by sampling every other frame. All frames are dual-channel ADF and BF STEM images. No data augmentation was used and the Adam optimiser with an initial learning rate of $10^{-5}$ was used to train the model. Noise2Void pixel masks were selected in a grid of pixels, with grid spacing equal to 24px. These grid-points were then modified/jittered by randomly translating them by up to two pixels in each direction (see Fig. \ref{fig: jitter}).

We also tested the performance of our modified Noise2Void model on new experimental data. We compared this to retraining of the model  (transfer training), where the previously trained model was used as a starting point and this was retrained on a subset of 615 images from the new experimental data. The original and retrained model was also compared to a completely new (randomly initialised) model trained exclusively on this new data (both using 615 frames from the total data set of 1230 frames and using all the new data) (section \ref{ssec: transfer-training}).

\subsection{Noise Metrics}

Values of PSNR, SSIM and N-RMSE were calculated using scikit-image's metrics module. SNR values were calculated by dividing the peak intensity (above the mean background) by the root-mean-square background noise.

\section{Data Availability}

Experimental videos, simulated images, model weights and denoised output is available upon reasonable request.

\section{Code Availability}

Code for training and applying the Noise2Void models, and for generating the simulations, is made available at \url{https://github.com/wilot/N2V-for-Au-GLC}.

\section{Author Information}

W.T. implemented the Noise2Void architecture, its modifications, performed the training, simulations, comparisons and timings. S.S.-A. and N.C. fabricated the graphene liquid cell samples. R.C. and S.S.-A. performed the TEM imaging. S.J.H. and R.G. supervised this research. W.T. wrote this manuscript and S.S.-A., N.C. and S.J.H. contributed to writing and editing this manuscript.

\section{Acknowledgements}

The authors acknowledge funding from the European Research Council (ERC) under the European Union’s Horizon 2020 research and innovation programme (Grant ERC-2016-STG-EvoluTEM-715502 and QTWIST (no. 101001515). We also thank the Engineering and Physical Sciences Research Council (EPSRC) for funding under grants EP/Y024303, EP/S021531/1, EP/M010619/1, EP/V007033/1, EP/S030719/1, EP/V001914/1, EP/V036343/1 and EP/P009050/1 and also for the EPSRC Centre for Doctoral Training (CDT) Graphene-NOWNANO. TEM access was supported by the Henry Royce Institute for Advanced Materials, funded through EPSRC grants EP/R00661X/1, EP/S019367/1, EP/P025021/1 and EP/P025498/1. RVG. acknowledges funding from the European Quantum Flagship Project 2DSIPC (no. 820378). We thank Diamond Light Source for access and support in use of the electron Physical Science Imaging Centre (Instrument E02 and proposal numbers MG33252 and MG35552) that contributed to the results presented here.

\section{Competing Interests}

All authors declare no financial or non-financial competing interests.

 \bibliographystyle{elsarticle-num} 
 \bibliography{cas-refs}






\clearpage

\section{Supplementary Information}

\setcounter{figure}{0}
\renewcommand{\figurename}{Fig.}
\renewcommand{\thefigure}{S\arabic{figure}}
\renewcommand{\thetable}{S\arabic{table}}

\begin{figure}[H]
    \centering
    \includegraphics[width=\linewidth]{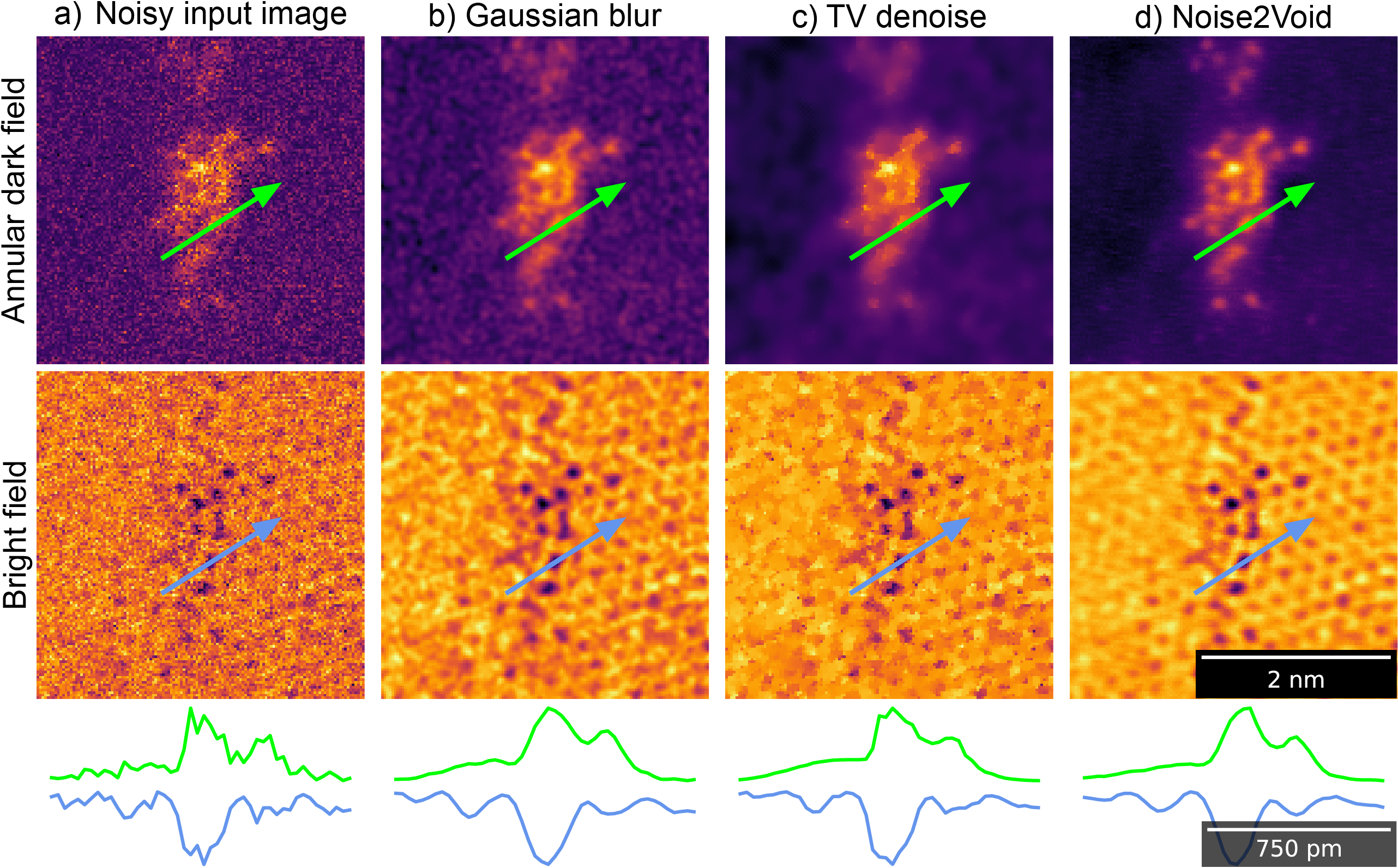}
    \caption{Demonstration of Noise2Void denoising performance on the experimental graphene liquid cell data and comparison to standard denoising methods. The line-profiles are taken at the same position for all data, (indicated by the green and blue arrows on ADF and BF images respectively) with a width of 1 px. From left to right the images compare the same frame a) from the original experimental data used as input for all the denoisers, b) after Gaussian denoising, c) after denoising by the total variation technique and d) after denoising by our modified Noise2Void approach. The square-root of pixel intensities are displayed for the ADF channel, though the BF images and all line-scans below are plotted linearly. Line profiles compare the intensity profile through two nearby Au adatoms separated by 0.34nm. The ADF profiles show that TV denoising is least able to recover the intensity minimum between the two adatoms, while Noise2Void recovers this well making the atoms are easier to distinguish as separate with the Noise2Void denoiser.}
    \label{fig: si-line-profile}
\end{figure}

\begin{figure}[H]
    \centering
    \includegraphics[width=\textwidth]{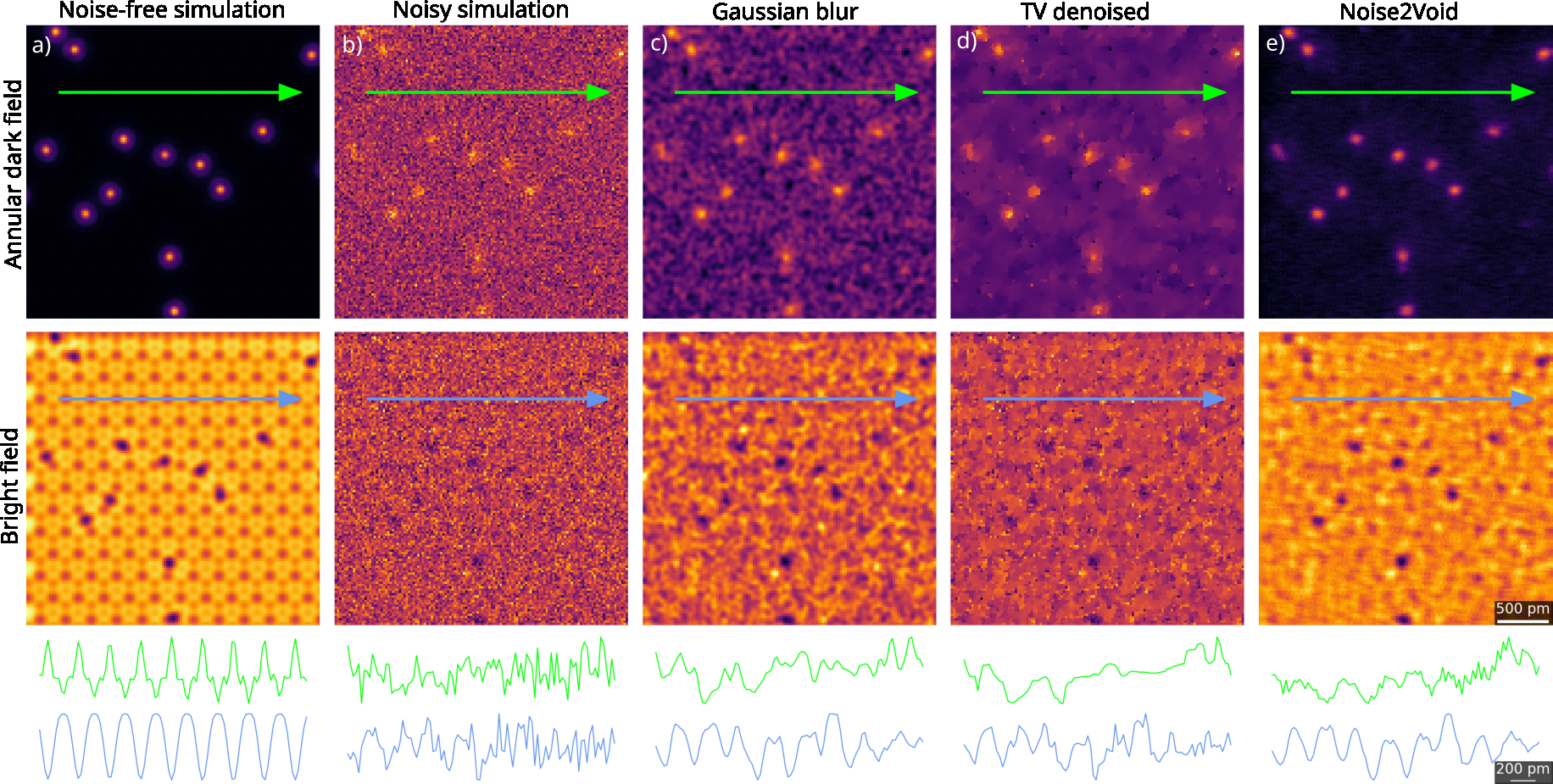}
    \caption{Denoising performance of the three denoising techniques compared, applied to a simulated liquid-cell frame. Frames are dual-channel annular dark-field (upper row) and bright-field (lower row). (a,b) are the noise free and noisy image simulations respectively while c-e show the simulation denoised by (c) Gaussian blurring, (d) total variation denoising, and (e) by our Noise2Void approach. Below each image is plotted an intensity line profile extracted with a summation width of 1px. Scale bar in a is the same for all panels.}
    \label{fig: sim_denoised_comparison-lattice-lprof}
\end{figure}

\begin{figure}[H]
    \centering
    \includegraphics[width=\linewidth]{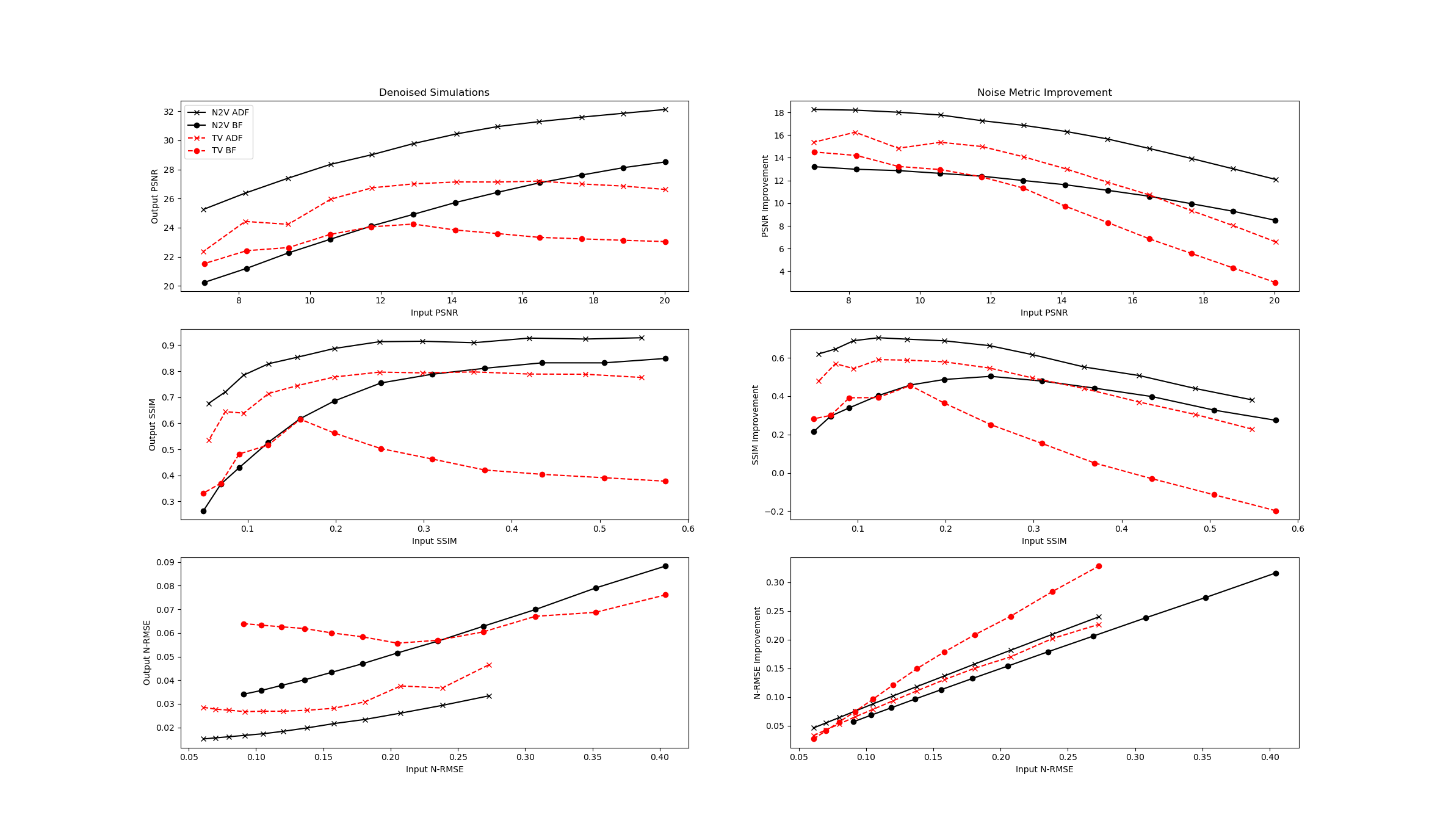}
    \caption{Noise2Void and TV denoising performance as a function of input noise. The same noise-free simulated image is used for all noisy inputs.}
    \label{fig: si-denoiser-performance-profile}
\end{figure}

\begin{figure}[H]
    \centering
    \includegraphics[width=\textwidth]{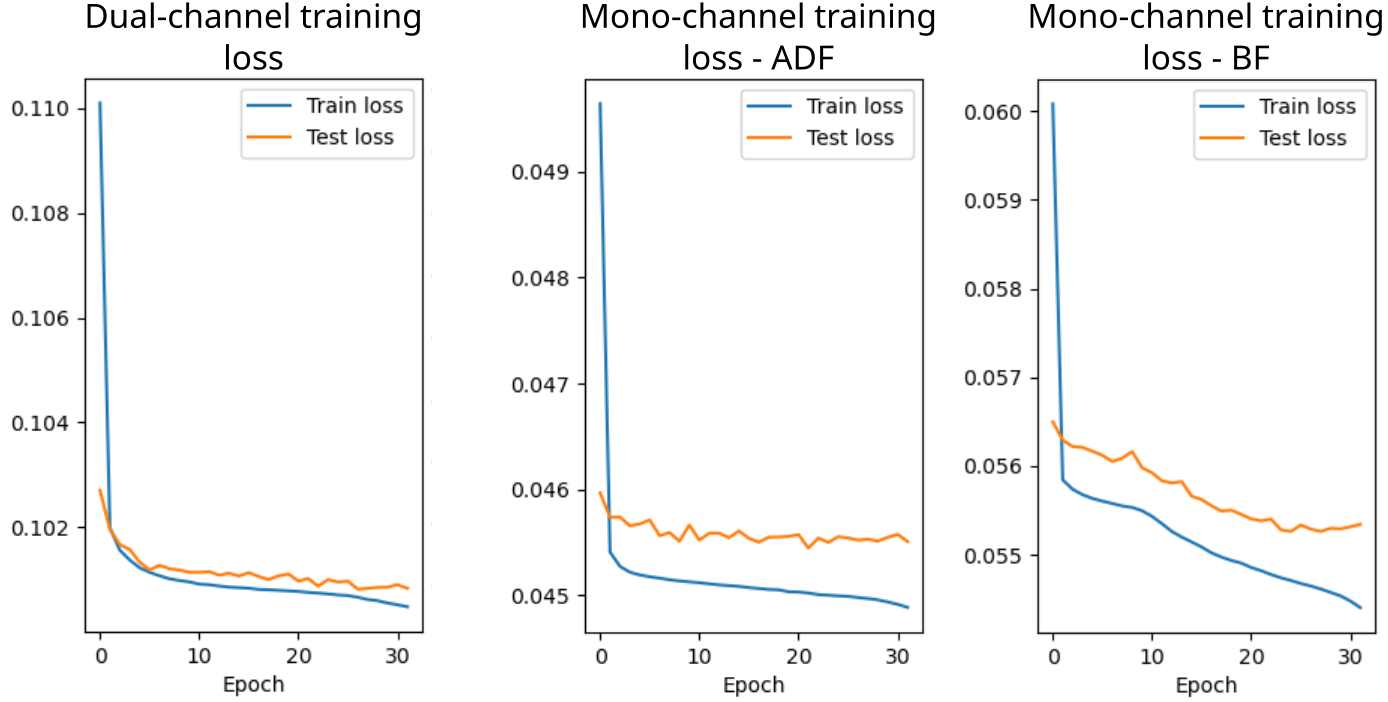}
    \caption{Demonstration of the training losses as a function of the number of epochs for the dual-channel network (used in Figs. \ref{fig: jitter}-\ref{fig: line-profile-cropped}) compared to a Noise2Void denoiser created using two seperate mono-channel UNets each trained on a single channel (ADF of BF images). These models were all trained for 32 epochs, with a mean-square-error loss (squared L2 norm) and an Adam optimiser with initial rate of $10^{-5}$. Values for the dual channel training loss are higher than the mono-channels as these are the sum of BF and ADF values.}
    \label{fig: training-losses}
\end{figure}

\begin{figure}[H]
    \centering
    \includegraphics[width=\linewidth]{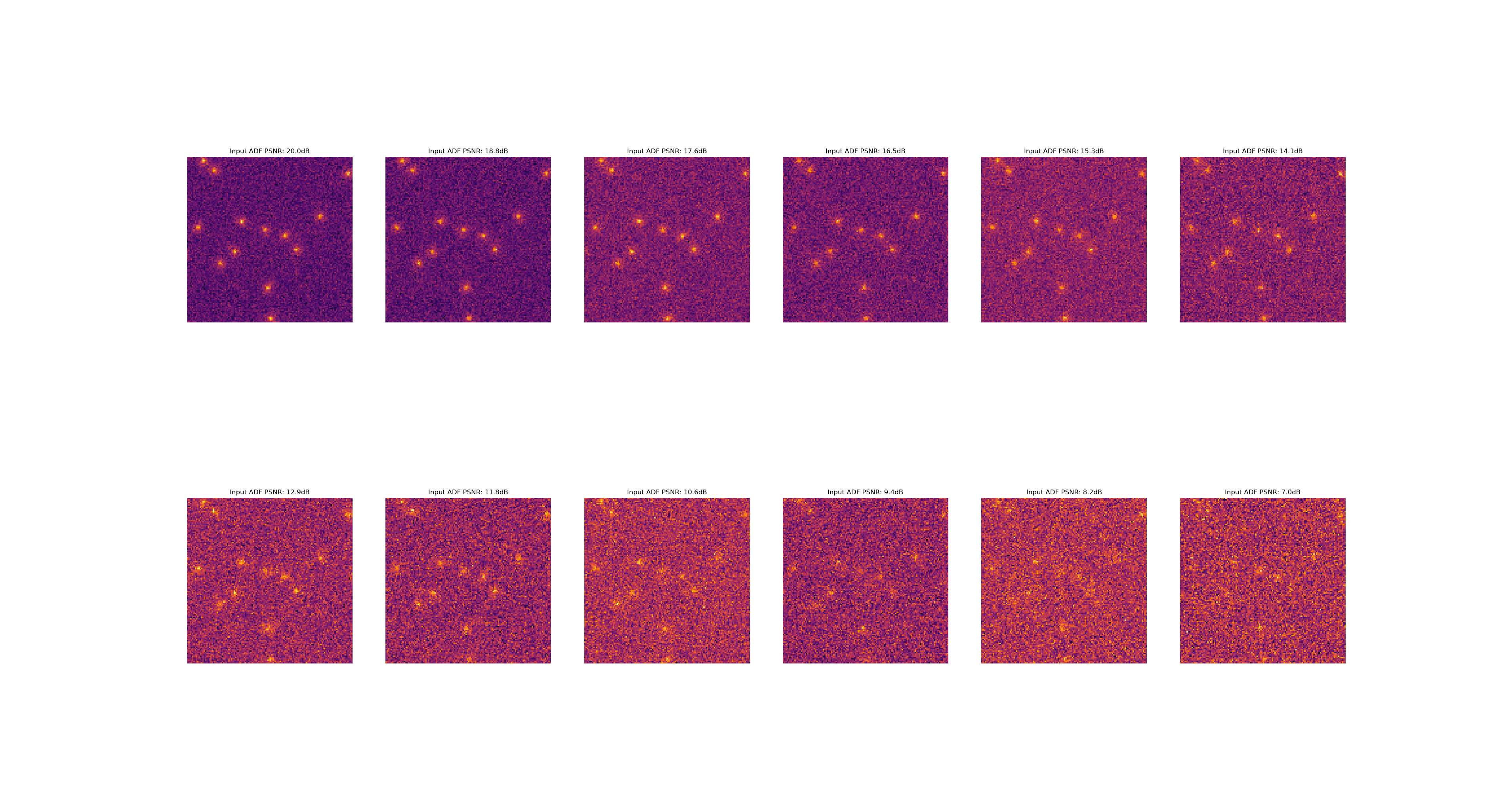}
    \caption{Input ADF images with increasing amounts of AGWN. These inputs correspond to the values plotted in Fig. \ref{fig: denoiser-performance-profile}.}
    \label{fig: denoiser-performance-profile-inputs}
\end{figure}

\begin{figure}[H]
    \centering
    \includegraphics[width=\linewidth]{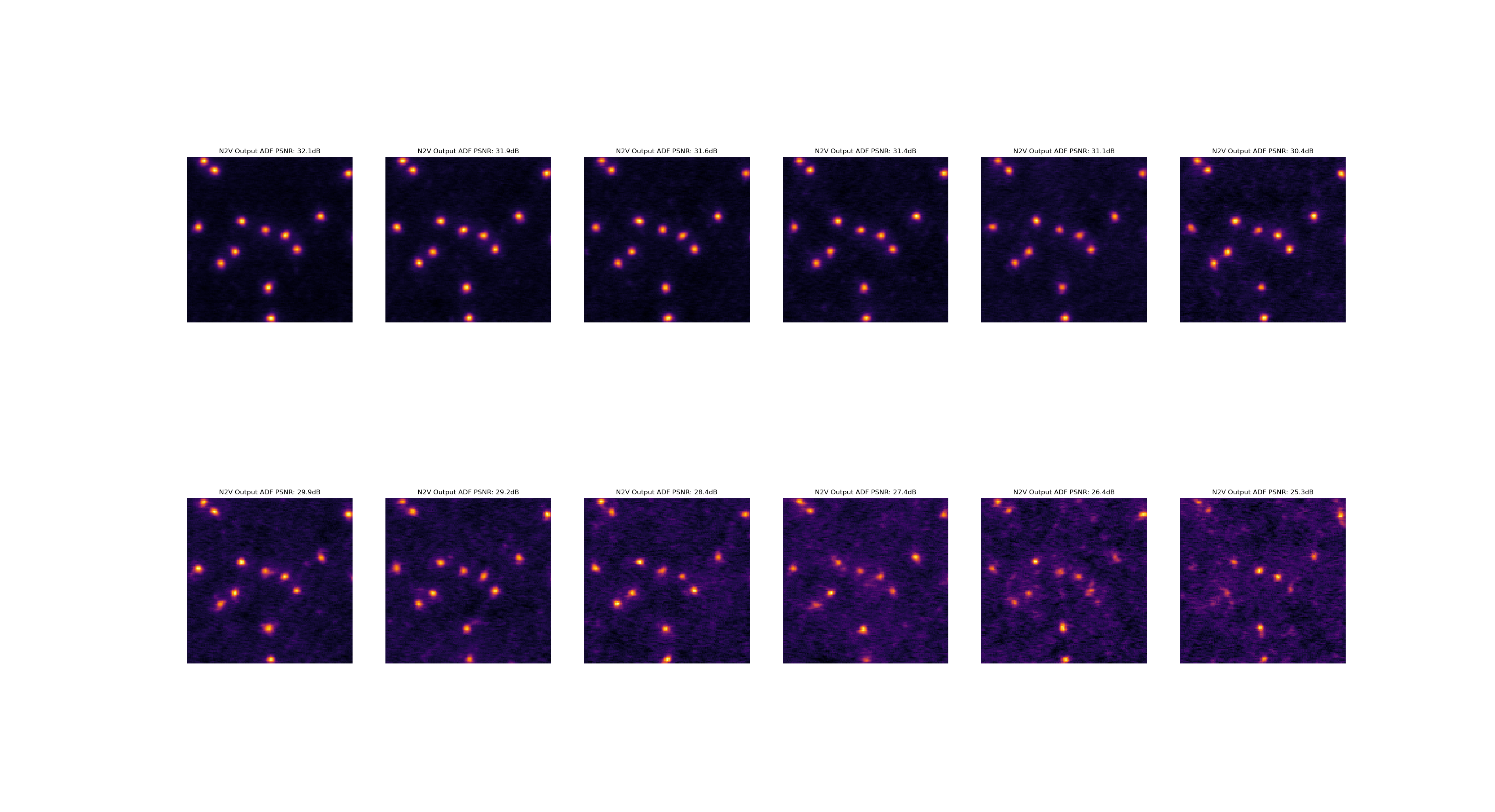}
    \caption{Noisy simulated images denoised with Noise2Void. The inputs are shown in SI Fig. \ref{fig: denoiser-performance-profile-inputs}.}
    \label{fig: denoiser-performance-profile-n2v-outputs}
\end{figure}

\begin{figure}[H]
    \centering
    \includegraphics[width=\linewidth]{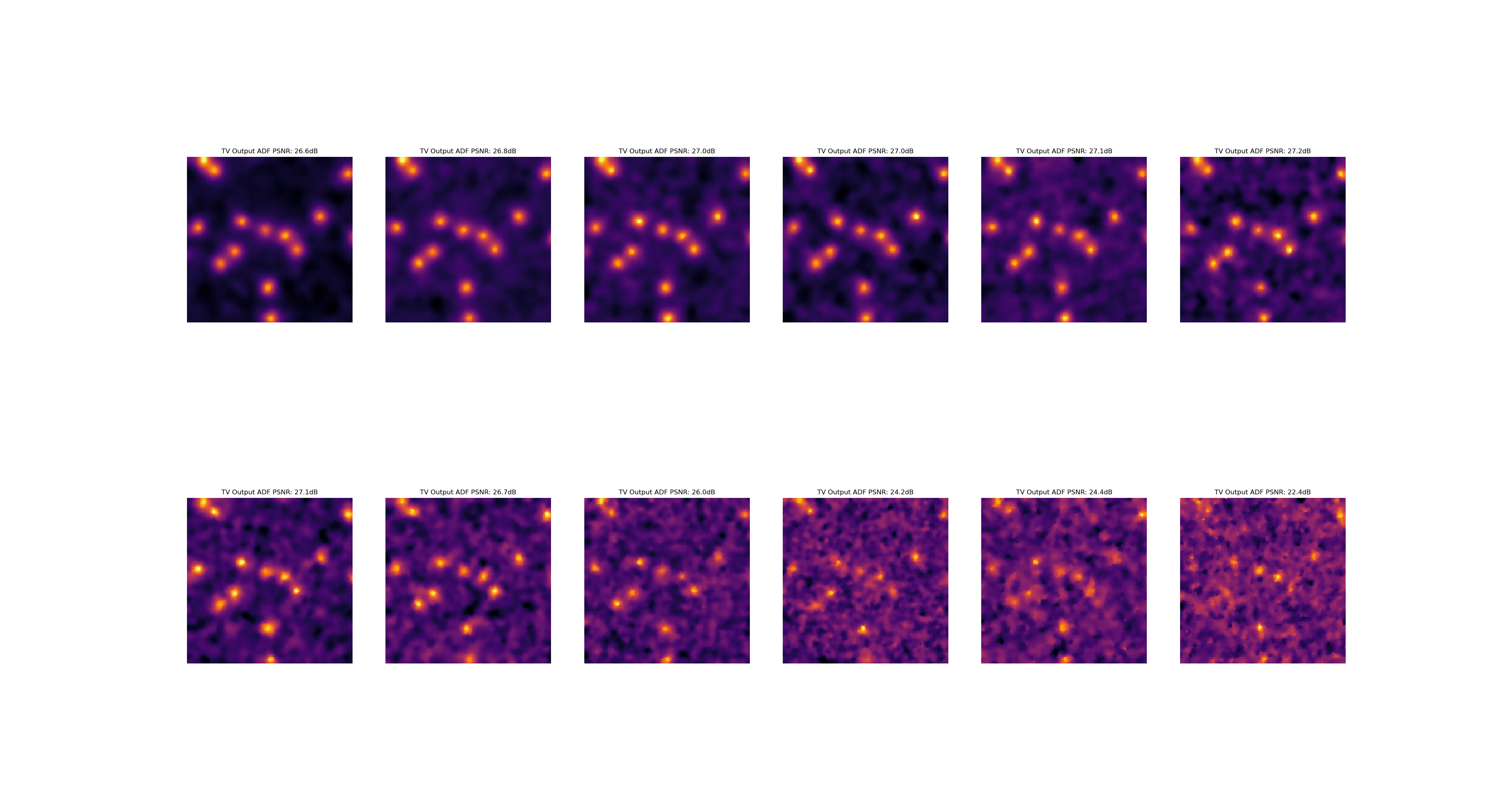}
    \caption{Noisy simulated images denoised with TV. The inputs are shown in SI Fig. \ref{fig: denoiser-performance-profile-inputs}.}
    \label{fig: denoiser-performance-profile-tv-outputs}
\end{figure}

\begin{table}[H]
    \centering
    \begin{tabular}{c||c|c|c|c}
        \hline
        ADF & Input & Gauss. & TV & N2V \\
        \hline
        PSNR (dB) & 7 & 18 & 22 & \textbf{23}\\
        SSIM & 0.026 & 0.104 & 0.223 & \textbf{0.360} \\
        N-RMSE & 0.27 & 0.08 & 0.05 & \textbf{0.04} \\
        \hline
    \end{tabular}
    
    \vspace{3mm}
    
    \begin{tabular}{c||c|c|c|c}
        \hline
        BF & Input & Gauss. & TV & N2V \\
        \hline
        PSNR (dB) & 6 & 17 & 20 & \textbf{21} \\
        SSIM & 0.021 & 0.116 & 0.132 & \textbf{0.185} \\
        N-RMSE & 0.46 & 0.13 & 0.09 & \textbf{0.08} \\
        \hline
    \end{tabular}
    \caption{Evaluation of the denoising performances of a simple Gaussian blur, TV denoising and Noise2Void. Shows denoising metrics when each technique is applied to simulated noisy images, for each channel (annular dark-field and bright-field) of the image separately, as shown in Fig. \ref{fig: sim_denoised_comparison}. These metrics compare the similarity of each denoiser's output (and the noisy input) to the noise-free simulation. The best values are shown in \textbf{bold}.}
\label{tab: sim_metrics}
\end{table}

\appendix

\end{document}